\documentclass[12pt]{iopart}

\usepackage[utf8]{inputenc}
\usepackage{booktabs}
\usepackage{siunitx}
\usepackage[T1]{fontenc}
\usepackage{mathptmx,amsmath,etoolbox,xcolor,amsfonts,amssymb,graphicx,dcolumn,bm}
\newcommand{\bra}[1]{\ensuremath{\left\langle#1\right|}}
\newcommand{\ket}[1]{\ensuremath{\left|#1\right\rangle}}
\newcommand{\bracket}[2]{\ensuremath{\left\langle #1 \middle| #2 \right\rangle}}
\usepackage[colorlinks=true, allcolors=blue]{hyperref}  

\begin{document}

\title[]{Generalization of Balian-Brezin decomposition for exponentials with linear fermionic part}
\author[1]{M. A. Seifi Mirjafarlou${}^{1}$, A. Jafarizadeh${}^{2}$, M. A. Rajabpour${}^{1}$}
\address{${}^{1}$Instituto de Fisica, Universidade Federal Fluminense, Av. Gal. Milton Tavares de Souza s/n, Gragoatá, 24210-346, Niterói, RJ, Brazil\\
${}^2$School of Physics and Astronomy, University of Nottingham, University Park, Nottingham NG7 2RD, United Kingdom}
\ead{adelseify86@gmail.com}
\vspace{10pt}
\begin{indented}
\item[\today]
\end{indented}

\begin{abstract}
Fermionic Gaussian states have garnered considerable attention due to their intriguing properties, most notably Wick's theorem. Expanding upon the work of Balian and Brezin, who generalized properties of fermionic Gaussian operators and states, we further extend their findings to incorporate Gaussian operators with a linear component. Leveraging a technique introduced by Colpa, we streamline the analysis and present a comprehensive extension of the Balian-Brezin decomposition (BBD) to encompass exponentials involving linear terms. Furthermore, we introduce Gaussian states featuring a linear part and derive corresponding overlap formulas. Additionally, we generalize Wick's theorem to encompass scenarios involving linear terms, facilitating the expression of generic expectation values in relation to one and two-point correlation functions. We also provide a brief commentary on the applicability of the BB decomposition in addressing the BCH (Zassenhaus) formulas within the  $\mathfrak{so}(N)$ Lie algebra.
\end{abstract}

%
%
%
%
%
\maketitle

\section{Introduction}
The study of Gaussian states, which are exact eigenstates of free fermion Hamiltonians or approximations to the ground state of interacting fermions, dates back to the early days of quantum mechanics with the introduction of Hartree-Fock-Bogoliubov variational techniques (see for example \cite{szabo2012modern}). Gaussian states possess many interesting properties, with one of the most celebrated being Wick's theorem.

In a highly useful paper, Balian and Brezin generalized various interesting properties of Gaussian operators and states for both bosonic and fermionic cases \cite{balian1969nonunitary}. Treating Gaussian operators as non-unitary canonical transformations, one of their useful formulas is the decomposition of a general Gaussian operator into three Gaussian operators, now known as the Balian-Brezin decomposition (BBD). Additionally, they introduced a more general version of Wick's theorem.

While the bosonic BBD in their paper is quite general, allowing for a linear part, the linear part is omitted in the case of fermionic systems. This omission seems justified since isolated fermionic systems with a linear part are not physically realizable. However, when dealing with open systems, see  \cite{shtanko2021complexity} or when the physical degrees of freedom differ from the fermionic degrees of freedom, such as in spin systems that can be mapped to free fermions via the Jordan-Wigner transformation, the linear part naturally arises and needs to be treated carefully, see \cite{jafarizadeh2022entanglement} and references therein. It is worth mentioning that Gaussian states with a linear part may serve as a useful variational ansatz too \cite{fukutome1977so,henderson2024hartree}.
In this paper, we extend the results of Balian and Brezin to Gaussian operators with a linear part. To accomplish this, we employ a trick introduced by Colpa, which significantly simplifies the treatment \cite{colpa1979diagonalisation}. One of the significant findings of our study is the extension of the BB decomposition to include exponentials with a linear part. Additionally, we introduce the concept of a Gaussian state with a linear part and derive the corresponding overlap formulas. Furthermore, we demonstrate the generalization of Wick's theorem to this more comprehensive scenario by showcasing how one can express generic expectation values in terms of the one and  two-point correlation functions.

The paper is organized as follows: In Section \eqref{sec: BBD}, after a brief introduction to BBD, we provide a comment on the case where a certain matrix $\mathbf{T}_{22}$ is singular, rendering the BBD inapplicable. In the same section, we present an explicit Pfaffian formula for the overlap of two general Gaussian states and also formulas for the correlation functions and the general Wick's theorem. While many of the formulas presented in this section have previously appeared in the literature, we provide a fresh presentation in order to establish a solid foundation for the subsequent sections. Then, in Section \eqref{sec: BBD with linear part}, we generalize the BB decomposition to exponentials with a linear part and derive general formulas for the overlaps. We also discuss how these formulas can be understood from the perspective of the Baker-Campbell-Hausdorff (BCH) formulas of the Lie algebra $\mathfrak{so}(N)$ (see for example \cite{zhang1990hsuan,gilmore2006lie,oeckl2014coherent}). Finally, in Section \eqref{sec: Generalized Wick}, we extend the generalized Wick formula of Balian-Brezin to cases that involve a linear part. Section \eqref{sec: Conclusions} concludes the paper.
In the appendix, we illustrate the steps of the introduced explicit methods through numerous examples. These examples serve to demonstrate how one can apply our main results, which are presented in equations \eqref{eq: generalized BBD}, \eqref{eq: generalized BCH}, \eqref{eq: generalized overlap}, \eqref{generalized Wick} and \eqref{eq: example of wick theorem}.

\section{Balian-Brezin decomposition: A review and a mini generalization}\label{sec: BBD}
In this section, we will first review the Balian-Brezin decomposition (BBD) and then extend it to the cases that certain matrices do not have inverses. Then we provide a general Pfaffian formula for the overlaps which can be also used to calculate all the correlation functions.

\subsection{Review}\label{subsec: review}
Consider the following operator:
\begin{equation}\label{eq: F}
\begin{split}
\mathcal{F}_{\mathbf{M}}=e^{ \frac{1}{2} \begin{pmatrix} \mathbf{c}^{\dagger} & \mathbf{c}\\ \end{pmatrix} \mathbf{M} \begin{pmatrix} \mathbf{c} \\ \mathbf{c}^{\dagger}  \end{pmatrix}},
\end{split}
\end{equation}
where $\left(\mathbf{c}^{\dagger}, \mathbf{c}\right)=\left(c_1^{\dagger},c_2^{\dagger}, ... , c_L^{\dagger},c_1,c_2, ... , c_L\right)$. Without losing generality, it is a requirement that $\mathbf{J} .\mathbf{M} $ be an antisymmetric matrix, where the $\mathbf{J}$ matrix is defined as follows:
\begin{equation}
\begin{split}
\mathbf{J} = \begin{pmatrix} \mathbf{0} & \mathbf{I} \\ \mathbf{I} & \mathbf{0} \end{pmatrix}.
\end{split}
\end{equation}
Using the Baker-Campbell-Hausdorff formula, one can show that \cite{balian1969nonunitary}:
\begin{equation}\label{eq: BCH for F}
\begin{split}
\mathcal{F}_{\mathbf{M}}^{-1}\begin{pmatrix} \mathbf{c} \\ \mathbf{c}^{\dagger}  \end{pmatrix} \mathcal{F}_{\mathbf{M}}=\mathbf{T} \begin{pmatrix} \mathbf{c} \\ \mathbf{c}^{\dagger}  \end{pmatrix}, \ \ \ \ \mathbf{T}=e^{\mathbf{M}},
\end{split}
\end{equation}
where $\mathbf{T}$ matrix satisfy the following property:
\begin{equation}
\begin{split}
\mathbf{T} \mathbf{J} \mathbf{T}^T=\mathbf{J}, \ \ \ \ \ \ \ \ \ \ \ \ \mathbf{T} = \begin{pmatrix} \mathbf{T_{11}} & \mathbf{T_{12}} \\ \mathbf{T_{21}} & \mathbf{T_{22}} \end{pmatrix}.
\end{split}
\end{equation}
This implies that the canonical fermionic anticommutation relations are preserved by the transformation \eqref{eq: BCH for F} (which is equivalent to saying that the transformation $\mathcal{F}_{\mathbf{M}}$ is canonical). In the following discussion, we will study the concept of canonical decomposition and examine various scenarios where submatrices of the $\mathbf{T}$ matrix are either invertible or non-invertible. In general, when two transformations $\mathcal{F}_{\mathbf{M}_1}$ and $\mathcal{F}_{\mathbf{M}_2}$ of the type \eqref{eq: F} are multiplied together, the resulting product $\mathcal{F}_{\mathbf{M}}$ can also be expressed as the exponential of a quadratic form as follows:
\begin{equation}\label{eq: BCH for F1 F2}
\begin{split}
e^{ \frac{1}{2} \begin{pmatrix} \mathbf{c}^{\dagger} & \mathbf{c}\\ \end{pmatrix} \mathbf{M}_1 \begin{pmatrix} \mathbf{c} \\ \mathbf{c}^{\dagger}  \end{pmatrix}} e^{ \frac{1}{2} \begin{pmatrix} \mathbf{c}^{\dagger} & \mathbf{c}\\ \end{pmatrix} \mathbf{M}_2 \begin{pmatrix} \mathbf{c} \\ \mathbf{c}^{\dagger}  \end{pmatrix}}=e^{ \frac{1}{2} \begin{pmatrix} \mathbf{c}^{\dagger} & \mathbf{c}\\ \end{pmatrix} \mathbf{M} \begin{pmatrix} \mathbf{c} \\ \mathbf{c}^{\dagger}  \end{pmatrix}},
\end{split}
\end{equation}
where:
\begin{equation} \label{eq: joint exponential}
\begin{split}
e^{\mathbf{M}_1}e^{\mathbf{M}_2}=e^{\mathbf{M}} \ \ \ \text{or} \ \ \mathbf{T}^{(1)}\mathbf{T}^{(2)}=\mathbf{T}.
\end{split}
\end{equation}
Using equation \eqref{eq: BCH for F1 F2}, one can apply the Balian-Brezin decomposition (BBD) to decompose the transformation $\mathcal{F}_{\mathbf{M}}$ as the product of three transformations $\mathcal{F}_{\mathbf{M}_1}$, $\mathcal{F}_{\mathbf{M}_2}$, and $\mathcal{F}_{\mathbf{M}_3}$ such that each one has just one of the terms in $\{cc,cc^{\dagger},c^{\dagger}c^{\dagger}\}$. This relation is widely used in the study of free fermion systems and has numerous applications, including the investigation of entanglement entropy \cite{fagotti2010entanglement}, full counting statistics \cite{najafi2017full}, and return amplitude after a quantum quench to name a few. There are two possible BBDs depending on which block of $\mathbf{T}$ is invertible. Note that the absence of an inverse in both cases indicates that the respective decomposition cannot be carried out.\\
\begin{itemize}
    \item We start by assuming that the $\mathbf{T}_{22}$ matrix is invertible. In this case, the Schur decomposition, also known as the LDU decomposition, would take the following form:
    \begin{subequations}
    \begin{equation}
\begin{split}
&\mathbf{T}=\begin{pmatrix} \mathbf{T}_{11} & \mathbf{T}_{12}\\ \mathbf{T}_{21} & \mathbf{T}_{22} \end{pmatrix} \equiv \begin{pmatrix} \mathbf{I} & \mathbf{T}_{12}\mathbf{T}_{22}^{-1} \\ \mathbf{0} & \mathbf{I} \end{pmatrix}\begin{pmatrix} \mathbf{T}_{11}-\mathbf{T}_{12}\mathbf{T}_{22}^{-1}\mathbf{T}_{21} & \mathbf{0}\\ \mathbf{0} & \mathbf{T}_{22}\end{pmatrix}\begin{pmatrix} \mathbf{I} & \mathbf{0} \\ \mathbf{T}_{22}^{-1}\mathbf{T}_{21} & \mathbf{I} \end{pmatrix},\\
\end{split}
\end{equation}
    \begin{equation}\label{eq: T_22 XYZ}
\begin{split}
& \mathbf{X}=\mathbf{T}_{12} (\mathbf{T}_{22})^{-1}, \ \ \ \ \ \mathbf{Z}=(\mathbf{T}_{22})^{-1}\mathbf{T}_{21},  \ \ \ \ \ e^{\mathbf{-Y}}=\mathbf{T}_{22}^T.
\end{split}
\end{equation}
\end{subequations}
The Balian-Brezin decomposition formula \cite{balian1969nonunitary} can be derived using the definitions provided above, as follows:
\begin{equation}\label{eq:BBD1}
\begin{split}
e^{ \frac{1}{2} \begin{pmatrix} \mathbf{c}^{\dagger} & \mathbf{c}\\ \end{pmatrix} \mathbf{M} \begin{pmatrix} \mathbf{c} \\ \mathbf{c}^{\dagger}  \end{pmatrix}} =e^{\frac{1}{2}\mathbf{c}^{\dagger} \mathbf{X}\mathbf{c}^{\dagger}} e^{\mathbf{c}^{\dagger} \mathbf{Y}\mathbf{c}-\frac{1}{2} Tr\mathbf{Y}}e^{\frac{1}{2}\mathbf{c} \mathbf{Z}\mathbf{c}}.
\end{split}
\end{equation}
\item If the $\mathbf{T_{11}}$ matrix is invertible, then we also have:
\begin{subequations}
\begin{equation}
\begin{split}
&\mathbf{T}=\begin{pmatrix} \mathbf{T}_{11} & \mathbf{T}_{12}\\ \mathbf{T}_{21} & \mathbf{T}_{22} \end{pmatrix} \equiv \begin{pmatrix} \mathbf{I} & \mathbf{0} \\ \mathbf{T}_{21}\mathbf{T}_{11}^{-1} & \mathbf{I} \end{pmatrix}\begin{pmatrix} \mathbf{T}_{11} & \mathbf{0}\\ \mathbf{0} & \mathbf{T}_{22}-\mathbf{T}_{21}\mathbf{T}_{11}^{-1}\mathbf{T}_{12} \end{pmatrix}\begin{pmatrix} \mathbf{I} & \mathbf{T}_{11}^{-1}\mathbf{T}_{12}\\ \mathbf{0} & \mathbf{I} \end{pmatrix},\\
\end{split}
\end{equation}
\begin{equation}
\begin{split}
&\mathbf{X}=\mathbf{T}_{21} (\mathbf{T}_{11})^{-1}, \ \ \ \ \ \mathbf{Z}=(\mathbf{T}_{11})^{-1}\mathbf{T}_{12} , \ \ \ \ \  e^{\mathbf{-Y}^T}=\mathbf{T}_{22}-\mathbf{T}_{21}\mathbf{T}_{11}^{-1}\mathbf{T}_{12}, \ \ \ \ \  e^{\mathbf{Y}}=\mathbf{T}_{11}.
\end{split}
\end{equation}
\end{subequations}
In the same way:
\begin{equation}\label{eq:BBD2}
\begin{split}
e^{ \frac{1}{2} \begin{pmatrix} \mathbf{c}^{\dagger} & \mathbf{c}\\ \end{pmatrix} \mathbf{M} \begin{pmatrix} \mathbf{c} \\ \mathbf{c}^{\dagger}  \end{pmatrix}} = e^{\frac{1}{2}\mathbf{c} \mathbf{X}\mathbf{c}} e^{\mathbf{c}^{\dagger} \mathbf{Y}\mathbf{c}-\frac{1}{2} Tr\mathbf{Y}} e^{\frac{1}{2}\mathbf{c}^{\dagger} \mathbf{Z}\mathbf{c}^{\dagger}}.
\end{split}
\end{equation}
\end{itemize}
Comparing with \eqref{eq:BBD1}, it becomes evident that the decomposed transformation represents a different form. The exponential term involving $c_ic_j$ couplings is now positioned on the left side, which can be interpreted as the replacement of all $c_j$ with $c_j^{\dagger}$ and vice versa. In the upcoming section, we will study the situations where the submatrices of $\mathbf{T}$ do not have necessarily an inverse.

\subsection{A mini generalization}\label{subsec: mini generalization}

The exponential (\ref{eq: F}) offers multiple possibilities for decomposition. However, in practical applications, the focus is often on the Balian-Brezin decomposition (BBD), which yields an exponential "normal ordered" form. Nevertheless, depending on the specific action of the operator or the non-invertibility of the matrix $\mathbf{T}_{22}$, alternative decompositions may be required. In this section, we discuss systematic approaches for studying such decompositions.

In general, when the submatrices of the $\mathbf{T}$ matrix are non-invertible or when a different type of decomposition is desired, one can opt for a canonical transformation, referred to as a "canonical permutation"(CP) from this point onward. Through this transformation, the applicability of the BBD can be restored.
In other words we write the \eqref{eq: F} as follows:
\begin{equation}
\begin{split}
e^{ \frac{1}{2} \begin{pmatrix} \mathbf{c}^{\dagger} & \mathbf{c}\\ \end{pmatrix} \boldsymbol{\Pi}^2_{j} \mathbf{M} \boldsymbol{\Pi}^2_{j} \begin{pmatrix} \mathbf{c} \\ \mathbf{c}^{\dagger}  \end{pmatrix}}=e^{ \frac{1}{2} \begin{pmatrix} \tilde{\mathbf{c}}^{\dagger} & \tilde{\mathbf{c}}\\ \end{pmatrix} \tilde{\mathbf{M}} \begin{pmatrix} \tilde{\mathbf{c}} \\ \tilde{\mathbf{c}}^{\dagger}  \end{pmatrix}}, \ \ \ \ \ \ \ \tilde{\mathbf{T}}=e^{\tilde{\mathbf{M}}},
\end{split}
\end{equation}
where the canonical permutation matrix can be defined as:
\begin{equation}
    \boldsymbol{\Pi}_{j}=\begin{pmatrix}
        \mathbf{I}_j & \mathbf{O}_j\\
        \mathbf{O}_j & \mathbf{I}_j\\
    \end{pmatrix},
    \quad\text{where}\quad
    \begin{cases}
                (\mathbf{I}_{j})_{nm}=\delta_{n,m}(1-\delta_{n,j}), & \\
                (\mathbf{O}_{j})_{nm}=\delta_{n,m}\delta_{n,j}, & \\
                 \end{cases}
\end{equation}
in which $\boldsymbol{\Pi}_{j}:\;c_j^\dagger\rightleftarrows c_{j}^{}$. The matrix we discussed earlier changes the arrangement of the operators $c_j^{\dagger}$ and $c_j$ within the set $\{c_1^{\dagger},..., c_L^{\dagger}, c_1,..., c_L\}$, this means that we can identify an original sub-matrix of elements in the matrix $\mathbf{T}$ which is non-singular. It's worth noting that this sub-matrix doesn't have to be a connected part of the matrix $\mathbf{T}$. As a result, the "canonical permutation" transformation becomes more general and powerful compared to the decompositions shown in the equations \eqref{eq:BBD1} and \eqref{eq:BBD2}. To provide a clearer understanding of the (CP) transformation, consider an example that highlights its relationship with the decompositions presented in the equation \eqref{eq:BBD2}. This example corresponds to a complete transformation where all the operators $c^{\dagger}$ are changed to $c$. When we apply the canonical permutation transformation to all the sites within the set $S=\{1,2,3,...,L\}$, the resulting expression becomes:
\begin{equation}
    \boldsymbol{\Pi}_{S}=
    \begin{pmatrix}
        \mathbf{0} & \mathbf{I}\\
        \mathbf{I} & \mathbf{0}\\
    \end{pmatrix}=\mathbf{J}.
\end{equation}
This permutation matrix is exactly what we need to obtain the expression \eqref{eq:BBD2}. In \ref{sec: an example}, we discuss an explicit example of size $L=3$ which helps to provide a clearer understanding of the two previous subsections. Note that generally, there is no assurance that a submatrix will possess an inverse following the CP transformation. Should an inverse be lacking, the associated decomposition becomes unfeasible. By conducting all the CP transformations and assessing the invertibility of each submatrix, one can identify all feasible decompositions.

\subsection{Overlap of arbitrary Gaussian states}\label{subsec: overlap}
The Balian-Brezin decomposition has useful applications, one of which is determining the explicit value of the overlap. Consider the state $\ket{\mathbf{I}}:=\ket{i_1,i_2,\cdots, i_L}$, where $i_n=0,1$, as one of the $2^L$ possible states on a configuration basis, then we define the state $\ket{\mathbf{M}_1(\mathbf{I})}$ as:
\begin{equation}
\begin{split}
\ket{\mathbf{M}_1(\mathbf{I})} :=e^{ \frac{1}{2} \begin{pmatrix} \mathbf{c}^{\dagger} & \mathbf{c}\\ \end{pmatrix} \mathbf{M}_1 \begin{pmatrix} \mathbf{c} \\ \mathbf{c}^{\dagger}  \end{pmatrix}} \ket{\mathbf{I}}.
\end{split}
\end{equation}
The goal is to calculate $\bracket{\mathbf{M}_2(\mathbf{J})}{\mathbf{M}_1(\mathbf{I})}$. Using equation (\ref{eq: BCH for F1 F2}) we have:
\begin{equation}
\begin{split}
\bracket{\mathbf{M}_2(\mathbf{J})}{\mathbf{M}_1(\mathbf{I})}=\bra{\mathbf{J}} \mathcal{F}_{\mathbf{M}} \ket{\mathbf{I}},
\end{split}
\end{equation}
where $e^{\mathbf{M}_2^{\dagger}}e^{\mathbf{M}_1}=e^{\mathbf{M}}$. Note that because of the parity number symmetry, one can immediately realize that the overlap is zero for those cases that $\mathbf{I}$ and $\mathbf{J}$ have different parities.  To calculate $\bra{\mathbf{J}} \mathcal{F}_{\mathbf{M}} \ket{\mathbf{I}}$ for the other cases, we first write this amplitude in the fermionic coherent basis $\ket{\boldsymbol{\xi}}$. It is defined as $c_j \ket{\boldsymbol{\xi}}=\xi_j \ket{\boldsymbol{\xi}}$, also as $\ket{\boldsymbol{\xi}}=e^{-\sum_{j=1}^{L} \xi_jc_j^{\dagger}}\ket{\mathbf{0}}$. In addition, we define $\mathbf{I}_0(\mathbf{I}_1)$ as the set of the sites with zero (one) fermion in increasing order. Then by using the Berezin integral over Grassmann variables, one can write:
\begin{subequations}
\begin{equation}
 \ket{\mathbf{I}}=(-1)^{\frac{|\mathbf{I}_1|(|\mathbf{I}_1|+1)}{2}} \int \prod_{j\in \mathbf{I}_1} d\xi_j \ket{\boldsymbol{\xi}(\mathbf{I})} ,\\
 \end{equation}
 \begin{equation}
 \bra{\mathbf{J}}=\int \prod_{k\in \mathbf{J}_1} d\delta_j \bra{\boldsymbol{\delta}(\mathbf{J})},
\end{equation}
\end{subequations}
where $|\mathbf{I}_1|$ is the size of $\mathbf{I}_1$, $\ket{\boldsymbol{\xi}(\mathbf{I})}$ and $\bra{\boldsymbol{\delta}(\mathbf{J})}$ are the coherent states in which we put $\xi_j=0$ for $j\in\mathbf{I}_0$ and $\delta_k=0$ for $k\in \mathbf{J}_0$. To write the $\bra{\mathbf{J}} \mathcal{F}_{\mathbf{M}} \ket{\mathbf{I}}$ in fermionic coherent basis we need to decompose the operator \eqref{eq: F} by using the BB decomposition and the completeness relation. Using the elementary Grassmann integral properties (see \cite{caracciolo2013algebraic}), we have:
\begin{equation}\label{eq: grassman integral}
\begin{split}
\bra{\mathbf{J}} \mathcal{F}_{\mathbf{M}} \ket{\mathbf{I}}=& (-1)^{\frac{|\mathbf{I}_1|(|\mathbf{I}_1|+1)}{2}} (-1)^{|\mathbf{I}_1||\mathbf{J}_1|} \int \mathbf{D}\boldsymbol{\xi}\mathbf{D}\boldsymbol{\gamma}\mathbf{D}\bar{\boldsymbol{\gamma}}\mathbf{D}\boldsymbol{\eta}\mathbf{D}\bar{\boldsymbol{\eta}}\mathbf{D}\bar{\boldsymbol{\delta}} \prod_{j\in\mathbf{J}_0} \bar{\delta}_j \prod_{k\in \mathbf{I}_0}\xi_k \left(\det \mathbf{T}_{22}\right)^{\frac{1}{2}} 
 \scriptsize e^{ \frac{1}{2} \begin{pmatrix} \bar{\boldsymbol{\delta}}  & \boldsymbol{\gamma} & \bar{\boldsymbol{\gamma}} & \boldsymbol{\eta} & \bar{\boldsymbol{\eta}} & \boldsymbol{\xi}   \\ \end{pmatrix} \mathcal{M} \begin{pmatrix} \bar{\boldsymbol{\delta}}  \\ \boldsymbol{\gamma} \\ \bar{\boldsymbol{\gamma}} \\ \boldsymbol{\eta} \\ \bar{\boldsymbol{\eta}} \\ \boldsymbol{\xi} \end{pmatrix}}\\
& =(-1)^{\frac{|\mathbf{I}_1|(|\mathbf{I}_1|+1)}{2}} (-1)^{|\mathbf{I}_1||\mathbf{J}_1|} \det[\mathbf{T}_{22}]^{\frac{1}{2}}\text{pf}[\mathcal{M}_{\mathbf{I}_0\mathbf{J}_0}],
\end{split}
\end{equation}
where
\begin{equation}\label{eq: M matrix}
\begin{split}
\mathcal{M}= \left(
\begin{array}{cccccc}
 \mathbf{X} & \mathbf{I} & \mathbf{0} & \mathbf{0} & \mathbf{0} & \mathbf{0} \\
 -\mathbf{I} & \mathbf{0} & \mathbf{I} & \mathbf{0} & \mathbf{0} & \mathbf{0} \\
 \mathbf{0} & -\mathbf{I} & \mathbf{0}& e^{\mathbf{Y}} & \mathbf{0} & \mathbf{0} \\
 \mathbf{0} & \mathbf{0} & -e^{\mathbf{Y}^T} & \mathbf{0} & \mathbf{I} & \mathbf{0} \\
 \mathbf{0} & \mathbf{0} & \mathbf{0} & -\mathbf{I} & \mathbf{0} & \mathbf{I} \\
 \mathbf{0} & \mathbf{0} & \mathbf{0} & \mathbf{0} & -\mathbf{I} & \mathbf{Z} \\
\end{array}
\right),
\end{split}
\end{equation}
and $\mathbf{X}$, $e^{\mathbf{Y}}$ and $\mathbf{Z}$ are the same as \eqref{eq: T_22 XYZ}, and $\mathcal{M}_{\mathbf{I}_0\mathbf{J}_0}$ is the matrix $\mathcal{M}$ in which we remove the rows $\mathbf{J}_0$ and $5L+\mathbf{I}_0$ and columns $\mathbf{J}_0$ and $5L+\mathbf{I}_0$. After some simplifications we have:
\begin{equation}\label{eq: reduced grassman integral}
\begin{split}
\bra{\mathbf{J}} \mathcal{F}_{\mathbf{M}} \ket{\mathbf{I}}= (-1)^{\frac{|\mathbf{I}_1|(|\mathbf{I}_1|+1)}{2}} (-1)^{|\mathbf{I}_1||\mathbf{J}_1|} \det[\mathbf{T}_{22}]^{\frac{1}{2}}\text{pf}[\mathcal{A}_{\mathbf{I}_0\mathbf{J}_0}],
\end{split}
\end{equation}
where
\begin{equation}\label{eq: A matrix}
\begin{split}
\mathcal{A}= \left(
\begin{array}{cc}
 \mathbf{X} & e^{\mathbf{Y}} \\
 -e^{\mathbf{Y}^T} & \mathbf{Z} \\
\end{array}
\right),
\end{split}
\end{equation}
and $\mathcal{A}_{\mathbf{I}_0\mathbf{J}_0}$ is the matrix $\mathcal{A}$ in which we remove the rows $\mathbf{J}_0$ and $L+\mathbf{I}_0$ and columns $\mathbf{J}_0$ and $L+\mathbf{I}_0$. A special case of the above equation first appeared in \cite{robledo2009sign} and then generalized in \cite{mizusaki2013grassmann}, see also \cite{najafi2019return}. If BB decomposition does not exist, in other words, if $\mathbf{T}_{22}$ does not have an inverse, we need to follow the procedure in the previous section and do canonical permutation transformation at sites $k\in S$ so that the new operator has a BB decomposition. After the first BB decomposition, we write all three exponentials in the original $(c,c^{\dagger})$ space. Each of the exponentials is again in the form \eqref{eq: F} and one should use again BB decomposition to further decompose them to new ones and continue this procedure up to the time that in the argument of each exponential, we have just the terms $c^{\dagger}c^{\dagger}$ or $c c$ or $c^{\dagger}c$. Then, one can write the operator in the fermionic coherent basis using multiple completeness relations and ultimately find the overlap. It is worth mentioning one can also find $|\bra{\mathbf{J}} \mathcal{F}_{\mathbf{M}} \ket{\mathbf{I}}|$ by the following simpler procedure: after the canonical permutation transformation that makes $\tilde{\mathbf{T}}_{22}$ invertible, one can just change the $\ket{\mathbf{I}}$ and $\bra{\mathbf{J}}$ to $\ket{\tilde{\mathbf{I}}}$ and $\bra{\tilde{\mathbf{J}}}$ so that $\tilde{\mathbf{I}}$ is the same as $\mathbf{I}$ except at sites $k\in S$, where we do the exchange $0\rightleftarrows1$, and we do the same with $\mathbf{J}$. Note that there is a sign ambiguity after the canonical transformation in this procedure. When $\mathbf{T}_{22}$ does not have an inverse sometimes it is possible to introduce a natural parameter $\epsilon$ in the matrix $\mathbf{M}$ such that the new $\mathbf{T}_{22}$ does have an inverse. Following the procedure by assuming $\epsilon$ is non-zero, one can find a result in which one can sometimes take the $\epsilon$ to zero without encountering any singularity. An example is shown in the \ref{subsec: Matrix elements}.

\subsection{Correlation functions}\label{subsec: Correlation functions}

In this subsection, we provide explicit formulas for the following $n$-point functions: 
\begin{equation}    
\langle A \rangle:=\bra{\mathbf{M}_2(\mathbf{J})}A\ket{\mathbf{M}_1(\mathbf{I})},
\end{equation}
where $A$ is an arbitrary product of $n$ creation and annihilation operators.
When $\ket{\mathbf{I}}$ and \ket{\mathbf{J}} are the vacuum states these correlations have been already calculated in \cite{balian1969nonunitary}, here we provide a pfaffian formula for arbitrary cases.
Because of Wick's theorem, we just need to calculate one and two-point functions. We note that since the parity of the $\ket{\mathbf{I}}$ and $\ket{\mathbf{J}}$ states can be different, the odd point functions are not necessarily zero. We start with one point function. When $\ket{\mathbf{I}}$ and $\ket{\mathbf{J}}$ have the same parity the expectation value is zero, otherwise one can write:
\begin{equation}    
\bra{\mathbf{J}}\mathcal{F}_{\mathbf{M}_2} c_i \mathcal{F}_{\mathbf{M}_1} \ket{\mathbf{I}}=\bra{\mathbf{J}}\mathcal{F}_{\mathbf{M}} \left(\mathbf{T}_{11}^{(1)} c + \mathbf{T}_{12}^{(1)}c^{\dagger}\right)_i \ket{\mathbf{I}},
\end{equation}
where $e^{\mathbf{M}_2^{\dagger}} e^{\mathbf{M}_1}=e^{\mathbf{M}}$, and $\mathbf{T}_{11}^{(1)}$ and $\mathbf{T}_{12}^{(1)}$ are the sub matrices of $\mathbf{T}^{(1)}=e^{\mathbf{M}_1}$. Since the $c$ and $c^{\dagger}$ act on the state $\ket{\mathbf{I}}=\ket{i_1,i_2,...,i_L}$ as follows:
\begin{equation}    
c_k\ket{\mathbf{I}}=\text{sign} (\mathbf{I}(k^-))\ket{\mathbf{I}(k^-)}, \ \ \ \ \ \ \ c_k^{\dagger}\ket{\mathbf{I}}=\text{sign} (\mathbf{I}(k^+))\ket{\mathbf{I}(k^+)},
\end{equation}
where 
\begin{equation}    
\text{sign} (\mathbf{I}(k^-))=\begin{cases}
    (-1)^{\sum_{p=1}^{k-1}i_p} \ \ \ \ \ \ \ \ i_k=1 \\
    0 \ \ \ \ \ \ \ \ \ \ \ \ \ \ \ \ \ \ \ \ \ i_k=0
\end{cases},  \text{sign} (\mathbf{I}(k^+))=\begin{cases}
      (-1)^{\sum_{p=1}^{k-1}i_p} \ \ \ \ \ \ \ \ i_k=0 \\
    0 \ \ \ \ \ \ \ \ \ \ \ \ \ \ \ \ \ \ \ \ \ i_k=1,
\end{cases}
\end{equation}
using the overlap formula \eqref{eq: reduced grassman integral} we have:
\begin{equation}
\begin{split}
\langle c_i \rangle =\det[\mathbf{T}_{22}]^{\frac{1}{2}}(-1)^{\frac{|\mathbf{I}_1|(|\mathbf{I}_1|-1)}{2}} (-1)^{|\mathbf{I}_1-1||\mathbf{J}_1|} &\Bigg[\text{sign} (\mathbf{I}(k^-)) \left(\mathbf{T}_{11}^{(1)}\right)_{ik} \text{pf}[\mathcal{A}_{\mathbf{I}_0(k^-)\mathbf{J}_0}]\\ & -
 \text{sign} (\mathbf{I}(k^+)) \left(\mathbf{T}_{12}^{(1)}\right)_{ik} \text{pf}[\mathcal{A}_{\mathbf{I}_0(k^+)\mathbf{J}_0}] \Bigg].
\end{split}
\end{equation}
Following a similar method, one can derive:
\begin{equation}
\begin{split}
\langle c_i^{\dagger} \rangle =\det[\mathbf{T}_{22}]^{\frac{1}{2}}(-1)^{\frac{|\mathbf{I}_1|(|\mathbf{I}_1|-1)}{2}} (-1)^{|\mathbf{I}_1-1||\mathbf{J}_1|} &\Bigg[\text{sign} (\mathbf{I}(k^-)) \left(\mathbf{T}_{21}^{(1)}\right)_{ik} \text{pf}[\mathcal{A}_{\mathbf{I}_0(k^-)\mathbf{J}_0}]\\& -
 \text{sign} (\mathbf{I}(k^+)) \left(\mathbf{T}_{22}^{(1)}\right)_{ik} \text{pf}[\mathcal{A}_{\mathbf{I}_0(k^+)\mathbf{J}_0}] \Bigg].
\end{split}
\end{equation}
Here we defined $\mathbf{I}(k^{\epsilon})$ as the set $\mathbf{I}$ minus (plus) the site $k$ for $\epsilon=-(+)$. $\mathbf{I}_0(k^{\epsilon})$ can be derived as before as the set of sites with zero fermions in $\mathbf{I}(k^{\epsilon})$.

The two-point functions can be calculated in the same way. Consider $\phi$ as either creation or annihilation operator with number $a,a^{\prime}=1(2)$ associated to $c(c^{\dagger})$. After some algebra we have:
\begin{equation}\label{two point correlation function}
\begin{split}
\langle \phi_i \phi_j \rangle =\det[\mathbf{T}_{22}]^{\frac{1}{2}}(-1)^{\frac{|\mathbf{I}_1|(|\mathbf{I}_1|+1)}{2}} (-1)^{|\mathbf{I}_1||\mathbf{J}_1|} &\Bigg[-\text{sign} (\mathbf{I}(k^-,l^-)) \left(\mathbf{T}_{a1}^{(1)}\right)_{ik} \left(\mathbf{T}_{a^{\prime}1}^{(1)}\right)_{jl} \text{pf}[\mathcal{A}_{\mathbf{I}_0(k^-,l^-)\mathbf{J}_0}]\\ &+ 
 \text{sign} (\mathbf{I}(k^-,l^+)) \left(\mathbf{T}_{a1}^{(1)}\right)_{ik} \left(\mathbf{T}_{a^{\prime}2}^{(1)}\right)_{jl} \text{pf}[\mathcal{A}_{\mathbf{I}_0(k^-,l^+)\mathbf{J}_0}]\\ & + 
 \text{sign} (\mathbf{I}(k^+,l^-)) \left(\mathbf{T}_{a2}^{(1)}\right)_{ik} \left(\mathbf{T}_{a^{\prime}1}^{(1)}\right)_{jl} \text{pf}[\mathcal{A}_{\mathbf{I}_0(k^+,l^-)\mathbf{J}_0}]\\ & -
 \text{sign} (\mathbf{I}(k^+,l^+)) \left(\mathbf{T}_{a2}^{(1)}\right)_{ik} \left(\mathbf{T}_{a^{\prime}2}^{(1)}\right)_{jl} \text{pf}[\mathcal{A}_{\mathbf{I}_0(k^+,l^+)\mathbf{J}_0}] \Bigg].
\end{split}
\end{equation}
The set $\mathbf{I}(k^{\epsilon_1},l^{\epsilon_2})$ is defined as follows: if $\epsilon$ is negative (positive) we remove (add) the corresponding sites $k,l$ from (to) the set $\mathbf{I}$. The signs $\mathbf{I}(k^{\epsilon_1},l^{\epsilon_2})$ are defined in the table \ref{table4}.\\
\begin{table}[h]
\centering
\caption{\footnotesize The sign of each term in two-point correlation function in the equation \eqref{two point correlation function} }\label{table4}
\renewcommand{\arraystretch}{1.9}
\begin{tabular}{|c|c|}
    \hline
    {\footnotesize {}} &  \begin{tabular}{@{}c@{}} Sign \end{tabular} \\
    \hline
    \text{sign}($\mathbf{I}(k^-,l^-)$)  & \begin{tabular}{@{}c@{}} $\begin{cases}
        (-1)^{\sum_{p=k}^{l-1}i_p} \ \ \ \ \ \ \ \ \ \ \ \ \ \ \{k<l, i_k=1, i_l=1\}\\
        (-1)^{\sum_{p=l}^{k-1}i_p+1} \ \ \ \ \ \ \ \ \ \ \{k>l, i_k=1, i_l=1\}\\
        0 \ \ \ \ \ \ \ \ \ \ \ \ \ \ \ \ \ \ \ \ \ \ \ \ \ \ \ \ \  \text{otherwise}
    \end{cases}$ \end{tabular} \\
    \hline
    \text{sign}($\mathbf{I}(k^-,l^+)$)  & \begin{tabular}{@{}c@{}} $\begin{cases}
        (-1)^{\sum_{p=k}^{l-1}i_p} \ \ \ \ \ \ \ \ \ \ \ \ \ \ \ \{k<l, i_k=1, i_l=0\}\\
        (-1)^{\sum_{p=l}^{k-1}i_p+1} \ \ \ \ \ \ \ \ \ \ \ \{k>l, i_k=1, i_l=0\}\\
        1 \ \ \ \ \ \ \ \ \ \ \ \ \ \ \ \ \ \ \ \ \ \ \ \ \ \ \ \ \ \ \{k=l,i_k=0\}\\
        0 \ \ \ \ \ \ \ \ \ \ \ \ \ \ \ \ \ \ \ \ \ \ \ \ \ \ \ \ \ \  \text{otherwise}
    \end{cases}$ \end{tabular} \\
    \hline
        \text{sign}($\mathbf{I}(k^+,l^-)$)  & \begin{tabular}{@{}c@{}} $\begin{cases}
        (-1)^{\sum_{p=k}^{l-1}i_p} \ \ \ \ \ \ \ \ \ \ \ \ \ \ \{k<l, i_k=0, i_l=1\}\\
        (-1)^{\sum_{p=l}^{k-1}i_p+1} \ \ \ \ \ \ \ \ \ \ \{k>l, i_k=0, i_l=1\}\\
        1 \ \ \ \ \ \ \ \ \ \ \ \ \ \ \ \ \ \ \ \ \ \ \ \ \ \ \ \ \ \  \{k=l,i_k=1\}\\
        0 \ \ \ \ \ \ \ \ \ \ \ \ \ \ \ \ \ \ \ \ \ \ \ \ \ \ \ \ \  \text{otherwise}
    \end{cases}$ \end{tabular} \\
    \hline
        \text{sign}($\mathbf{I}(k^+,l^+)$)  & \begin{tabular}{@{}c@{}} $\begin{cases}
        (-1)^{\sum_{p=k}^{l-1}i_p} \ \ \ \ \ \ \ \ \ \ \ \ \ \{k<l, i_k=0, i_l=0\}\\
        (-1)^{\sum_{p=l}^{k-1}i_p+1} \ \ \ \ \ \ \ \ \ \{k>l, i_k=0, i_l=0\}\\
        0 \ \ \ \ \ \ \ \ \ \ \ \ \ \ \ \ \ \ \ \ \ \ \ \ \ \ \ \  \text{otherwise}
    \end{cases}$ \end{tabular} \\
    \hline
\end{tabular}
\end{table}\\
Note that when $\ket{\mathbf{I}}$ and $\ket{\mathbf{J}}$ have similar (opposite) parities the odd (even) point functions are zero.
The higher-point functions can be calculated using the Wick's theorem. In the case of similar parities of $\ket{\mathbf{I}}$ and $\ket{\mathbf{J}}$ even point functions can be written as the products of two-point functions with appropriate signs. For example, one can write:
\begin{equation}
\begin{split}    
& \langle \phi_i \phi_j \phi_k \phi_l \rangle= \bracket{\mathbf{M}_2(\mathbf{J})}{\mathbf{M}_1(\mathbf{I})}^{-1} \bigg[\langle \phi_i \phi_j \rangle \langle \phi_k \phi_l \rangle -\langle \phi_i \phi_k \rangle \langle \phi_j \phi_l \rangle + \langle \phi_i \phi_l \rangle \langle \phi_j \phi_k \rangle\bigg].
\end{split}
\end{equation}
Note that for $2n$-point functions we need to divide the right side with $\bracket{\mathbf{M}_2(\mathbf{J})}{\mathbf{M}_1(\mathbf{I})}^{n/2}$.
The case of opposite parities and odd point functions is a bit subtle. To calculate these correlation functions we first write $\ket{\mathbf{I}}=\phi_k \ket{\mathbf{I}^{\prime}}$ so that $\ket{\mathbf{I}^{\prime}}$ and $\ket{\mathbf{J}}$ have the same parities. For example, consider $\phi_k=c_k$, then: 
\begin{equation}
\begin{split}    
\bra{\mathbf{M}_2(\mathbf{J})} A \ket{\mathbf{M}_1(\mathbf{I})} &=\bra{\mathbf{M}_2(\mathbf{J})} A(\mathbf{T}_{11}^{(-1)}.c+\mathbf{T}_{12}^{(-1)}.c^{\dagger})_k \ket{\mathbf{M}_1(\mathbf{I}^{\prime})}  \\&=\left(\mathbf{T}_{11}^{(-1)}\right)_{kj}\bra{\mathbf{M}_2(\mathbf{J})}  A c_j \ket{\mathbf{M}_1(\mathbf{I}^{\prime})} 
 +\left(\mathbf{T}_{12}^{(-1)}\right)_{kj} 
 \bra{\mathbf{M}_2(\mathbf{J})} A c_j^{\dagger})\ket{\mathbf{M}_1(\mathbf{I}^{\prime})},
\end{split}
\end{equation}
where $\mathbf{T}^{(-1)}=e^{-\mathbf{M}_1}$. One can now apply Wick's theorem to each one of the above expectation values. The generalization of Wick's theorem from the perspective of the fermionic coherent basis has been previously discussed in \cite{mizusaki2012new}, see also \cite{robledo2009sign} for the earlier study that shows the connection to the Pfaffian formulas. For a more recent discussion on off-diagonal Wick's theorem and other related methods, refer to \cite{porro2022off}.

\section{BB decomposition for exponentials with linear part}\label{sec: BBD with linear part}

In this section, we would like to decompose the following exponential
\begin{equation}
\begin{split}
\mathcal{F}_{(\mathbf{M,u,v})}=e^{ \frac{1}{2} \begin{pmatrix} \mathbf{c}^{\dagger} & \mathbf{c}\\ \end{pmatrix} \mathbf{M} \begin{pmatrix} \mathbf{c} \\ \mathbf{c}^{\dagger}  \end{pmatrix}+\mathbf{u}^{\dagger}\mathbf{c}^{\dagger}+\mathbf{v}^{T}\mathbf{c}}
\end{split}
\end{equation}
to the multiplication of various exponentials so that each exponential has just one of the terms in the set \{$c^{\dagger}c^{\dagger}$, $cc$, $c^{\dagger}c$, $c^{\dagger}$ or $c$\}.

\subsection{Ancilary site method}\label{subsec: ancillary site}

To handle the linear part we use the trick which was first suggested by Colpa \cite{colpa1979diagonalisation}. For an introduction to the method see the Appendix C. The idea is based on introducing an ancillary site and the following substitutions:
\begin{equation}\label{eq: ancillary site}
\begin{split}
c_j\rightarrow c_0^{\dagger}c_j-c_0c_j, \ \ \ \ \ \ \ \ \ c_j^{\dagger}\rightarrow c_j^{\dagger}c_0-c_j^{\dagger}c_0^{\dagger},
\end{split}
\end{equation}
which keeps the quadratic term unchanged, however, it changes the linear parts to quadratic form such that we have:
\begin{equation}\label{eq: F Mprime}
\begin{split}
\mathcal{F}_{\mathbf{M}^{\prime}}=e^{ \frac{1}{2} \begin{pmatrix} \mathbf{c}^{\dagger} & \mathbf{c}\\ \end{pmatrix} \mathbf{M}^{\prime}\begin{pmatrix} \mathbf{c} \\ \mathbf{c}^{\dagger}  \end{pmatrix}},
\end{split}
\end{equation}
where:
\begin{equation}\label{eq: new Mprime}
\begin{split}
\mathbf{M}^{\prime}= \begin{pmatrix} 0 & \vec{\mathbf{v}}^T& 0&  \vec{\mathbf{u}}^{\dagger} \\   \vec{\mathbf{u}}^{\ast} & \mathbf{M}_{11} & - \vec{\mathbf{u}}^{\ast} & \mathbf{M}_{12} \\ 0 & - \vec{\mathbf{v}}^{T} & 0 & - \vec{\mathbf{u}}^{\dagger} \\  \vec{\mathbf{v}} & \mathbf{M}_{21} & - \vec{\mathbf{v}} & \mathbf{M}_{22}\end{pmatrix}, \ \ \ \ \ \begin{pmatrix} \mathbf{c}^{\dagger} & \mathbf{c}\\ \end{pmatrix}= \begin{pmatrix}c_0^{\dagger} & c_1^{\dagger} & \cdots & c_L^{\dagger} & c_0 & c_1 & \cdots & c_L\\ \end{pmatrix}.
\end{split}
\end{equation}
Applying the equations \eqref{eq: BCH for F1 F2} and \eqref{eq: joint exponential} one can now write:
\begin{equation}
\mathcal{F}_{(\mathbf{M}_1,\mathbf{u}_1,\mathbf{v}_1)} \mathcal{F}_{(\mathbf{M}_2,\mathbf{u}_2,\mathbf{v}_2)}=\mathcal{F}_{(\mathbf{M},\mathbf{u},\mathbf{v})}, \ \ \ \ \ \ e^{\mathbf{M}_1^{\prime}} e^{\mathbf{M}_2^{\prime}}=e^{\mathbf{M}^{\prime}},
\end{equation}
where $\mathbf{M}_1^{\prime}$, $\mathbf{M}_2^{\prime}$ and $\mathbf{M}^{\prime}$ are built using equation \eqref{eq: new Mprime}. There are many possible decompositions for the exponential \eqref{eq: F Mprime}, however, the most useful one is the following decomposition:
\begin{equation}\label{eq: BBD Fnew}
\mathcal{F}_{(\mathbf{M,u,v})} =e^{\mathbf{q}^{\dagger}\mathbf{c}^{\dagger}}e^{\frac{1}{2}\mathbf{c}^{\dagger} \mathbf{X}\mathbf{c}^{\dagger}} e^{\mathbf{c}^{\dagger} \mathbf{Y}\mathbf{c}-\frac{1}{2} Tr\mathbf{Y}}e^{\frac{1}{2}\mathbf{c} \mathbf{Z}\mathbf{c}} e^{\mathbf{p}^T\mathbf{c}}.
\end{equation}
To find $\mathbf{X}$, $\mathbf{Y}$ and $\mathbf{Z}$ matrices and $\mathbf{p}$ and $\mathbf{q}$ vectors, we apply the Colpa trick also on the arguments of the exponentials of the right-hand side of the equation \eqref{eq: BBD Fnew}. Then the right side will be:
\begin{equation}
\begin{split}
e^{ \frac{1}{2} \begin{pmatrix} \mathbf{c}^{\dagger} & \mathbf{c}\\ \end{pmatrix} \mathbf{M}'_q\begin{pmatrix} \mathbf{c} \\ \mathbf{c}^{\dagger}  \end{pmatrix}} e^{ \frac{1}{2} \begin{pmatrix} \mathbf{c}^{\dagger} & \mathbf{c}\\ \end{pmatrix} \mathbf{M}'_X\begin{pmatrix} \mathbf{c} \\ \mathbf{c}^{\dagger}  \end{pmatrix}} e^{ \frac{1}{2} \begin{pmatrix} \mathbf{c}^{\dagger} & \mathbf{c}\\ \end{pmatrix} \mathbf{M}'_Y\begin{pmatrix} \mathbf{c} \\ \mathbf{c}^{\dagger}  \end{pmatrix}} e^{ \frac{1}{2} \begin{pmatrix} \mathbf{c}^{\dagger} & \mathbf{c}\\ \end{pmatrix} \mathbf{M}'_Z\begin{pmatrix} \mathbf{c} \\ \mathbf{c}^{\dagger}  \end{pmatrix}} e^{ \frac{1}{2} \begin{pmatrix} \mathbf{c}^{\dagger} & \mathbf{c}\\ \end{pmatrix} \mathbf{M}'_p\begin{pmatrix} \mathbf{c} \\ \mathbf{c}^{\dagger}  \end{pmatrix}},
\end{split}
\end{equation}
where:
\begin{equation}
\begin{split}
& \mathbf{M}'_q= \begin{pmatrix} 0 & \vec{\mathbf{0}}& 0&  \vec{\mathbf{q}}^{\dagger} \\ \vec{\mathbf{q}}^{\ast} & \mathbf{0} & -\vec{\mathbf{q}}^{\ast} & \mathbf{0} \\ 0 & \vec{\mathbf{0}} & 0 & -\vec{\mathbf{q}}^{\dagger} \\ \vec{\mathbf{0}} & \mathbf{0} & \vec{\mathbf{0}} & \mathbf{0}\end{pmatrix},  \ \ \ \ \ \mathbf{M}'_X= \begin{pmatrix} 0 & \vec{\mathbf{0}}& 0&  \vec{\mathbf{0}} \\ \vec{\mathbf{0}} & \mathbf{0} & \vec{\mathbf{0}} & \mathbf{X} \\ 0 & \vec{\mathbf{0}} & 0 & \vec{\mathbf{0}} \\ \vec{\mathbf{0}} & \mathbf{0} & \vec{\mathbf{0}} & \mathbf{0}\end{pmatrix}, \ \ \ \ \ \mathbf{M}'_Y= \begin{pmatrix} 0 & \vec{\mathbf{0}}& 0&  \vec{\mathbf{0}} \\ \vec{\mathbf{0}} & \mathbf{Y} & \vec{\mathbf{0}} & \mathbf{0} \\ 0 & \vec{\mathbf{0}} & 0 & \vec{\mathbf{0}} \\ \vec{\mathbf{0}} & \mathbf{0} & \vec{\mathbf{0}} & -\mathbf{Y}^T\end{pmatrix},\\
& \mathbf{M}'_Z= \begin{pmatrix} 0 & \vec{\mathbf{0}}& 0& \vec{\mathbf{0}} \\ \vec{\mathbf{0}} & \mathbf{0} & \vec{\mathbf{0}} & \mathbf{0} \\ 0 & \vec{\mathbf{0}} & 0 & \vec{\mathbf{0}} \\ \vec{\mathbf{0}} & \mathbf{Z} & \vec{\mathbf{0}} & \mathbf{0}\end{pmatrix}, \ \ \ \ \ \ \mathbf{M}'_p= \begin{pmatrix} 0 & \vec{\mathbf{p}}^T& 0&  \vec{\mathbf{0}} \\ \vec{\mathbf{0}} & \mathbf{0} & \vec{\mathbf{0}} & \mathbf{0} \\ 0 & -\vec{\mathbf{p}}^T & 0 & \vec{\mathbf{0}} \\ \vec{\mathbf{p}} & \mathbf{0} & -\vec{\mathbf{p}} & \mathbf{0}\end{pmatrix},
\end{split}
\end{equation}
and
\begin{equation}\label{eq: exponentials}
\begin{split}
&e^{\mathbf{M}'_q} e^{ \mathbf{M}'_X} e^{\mathbf{M}'_Y} e^{\mathbf{M}'_Z} e^{ \mathbf{M}'_p}=\\
& \scriptsize \begin{pmatrix} 1+\vec{\mathbf{q}}^{\dagger}e^{-\mathbf{Y}^T}\vec{\mathbf{p}} & \vec{\mathbf{p}}^T+\vec{\mathbf{q}}^{\dagger}e^{-\mathbf{Y}^T} \mathbf{Z}+\vec{\mathbf{q}}^{\dagger}e^{-\mathbf{Y}^T}\vec{\mathbf{p}}\vec{\mathbf{p}}^T& -\vec{\mathbf{q}}^{\dagger}e^{-\mathbf{Y}^T}\vec{\mathbf{p}} &  \vec{\mathbf{q}}^{\dagger}e^{-\mathbf{Y}^T} \\ \vec{\mathbf{q}}^{\ast}+\mathbf{X}e^{-\mathbf{Y}^T}\vec{\mathbf{p}}+ \vec{\mathbf{q}}^{\ast}\vec{\mathbf{q}}^{\dagger}e^{-\mathbf{Y}^T}\vec{\mathbf{p}} & 2\vec{\mathbf{q}}^{\ast}\vec{\mathbf{p}}^{T}+e^{\mathbf{Y}}+\mathbf{X}e^{-\mathbf{Y}^T}\mathbf{Z}+\vec{\mathbf{q}}^{\ast}\vec{\mathbf{q}}^{\dagger}e^{-\mathbf{Y}^T}\mathbf{Z}+\mathbf{X}e^{-\mathbf{Y}^T}\vec{\mathbf{p}}\vec{\mathbf{p}}^{T}+\vec{\mathbf{q}}^{\ast}\vec{\mathbf{q}}^{\dagger}e^{-\mathbf{Y}^T}\vec{\mathbf{p}}\vec{\mathbf{p}}^{T} & -\vec{\mathbf{q}}^{\ast}-\mathbf{X}e^{-\mathbf{Y}^T}\vec{\mathbf{p}}- \vec{\mathbf{q}}^{\ast}\vec{\mathbf{q}}^{\dagger}e^{-\mathbf{Y}^T}\vec{\mathbf{p}} & \mathbf{X}e^{-\mathbf{Y}^T}+\vec{\mathbf{q}}^{\ast}\vec{\mathbf{q}}^{\dagger} e^{-\mathbf{Y}^T} \\ -\vec{\mathbf{q}}^{\dagger}e^{-\mathbf{Y}^T}\vec{\mathbf{p}} & -\vec{\mathbf{p}}^T-\vec{\mathbf{q}}^{\dagger}e^{-\mathbf{Y}^T} \mathbf{Z}-\vec{\mathbf{q}}^{\dagger}e^{-\mathbf{Y}^T}\vec{\mathbf{p}}\vec{\mathbf{p}}^T & 1+\vec{\mathbf{q}}^{\dagger}e^{-\mathbf{Y}^T}\vec{\mathbf{p}} & -\vec{\mathbf{q}}^{\dagger}e^{-\mathbf{Y}^T} \\ e^{-\mathbf{Y}^T}\vec{\mathbf{p}} & e^{-\mathbf{Y}^T}\mathbf{Z}+e^{-\mathbf{Y}^T}\vec{\mathbf{p}}\vec{\mathbf{p}}^T & -e^{-\mathbf{Y}^T}\vec{\mathbf{p}} & e^{-\mathbf{Y}^T}\end{pmatrix}.
\end{split}
\end{equation}
On the other hand, we also have:
\begin{equation}\label{eq: T prime matrix}
\begin{split}
& \mathbf{T}'= e^{\mathbf{M}'}=\begin{pmatrix} 1+t_{11} & \vec{\mathbf{t}}_1^T& -t_{11}&  \vec{\mathbf{t}}_2^T \\ \vec{\mathbf{t}}_3 & \mathbf{T}_{11} & -\vec{\mathbf{t}}_3 & \mathbf{T}_{12} \\ -t_{11} & -\vec{\mathbf{t}}_1^T & 1+t_{11} & -\vec{\mathbf{t}}_2^T \\ \vec{\mathbf{t}}_4 & \mathbf{T}_{21} & -\vec{\mathbf{t}}_4 & \mathbf{T}_{22}\end{pmatrix}.
\end{split}
\end{equation} 
Putting the equations \eqref{eq: exponentials} and \eqref{eq: T prime matrix} equal, one can get the following equalities:
\begin{equation}\label{eq: generalized BBD}
\begin{split}
&e^{-\mathbf{Y}^T}=\mathbf{T}_{22}, \ \ \ \ \ \vec{\mathbf{p}}=\mathbf{T}_{22}^{-1}  \vec{\mathbf{t}}_4, \ \ \ \ \ \ \vec{\mathbf{q}}=\vec{\mathbf{t}}_2^T\mathbf{T}_{22},\\
& \mathbf{Z}=\mathbf{T}_{22}^{-1}\mathbf{T}_{21}-\vec{\mathbf{p}}\vec{\mathbf{p}}^T, \ \ \ \ \ \ \ \ \mathbf{X}=\mathbf{T}_{12}\mathbf{T}_{22}^{-1} - \vec{\mathbf{q}}^{\ast} \vec{\mathbf{q}}^{\dagger}.
\end{split}
\end{equation}
Note that when $\mathbf{u}=\mathbf{v}=\mathbf{0}$, we have the original BB decomposition. In \ref{sec: Example of decomposition of exponential with linear term} we show a simple example of the above procedure.

\subsection{BCH (Zassenhaus) formula for $\mathfrak{so}(2N+1)$}\label{subsec: BCH formula}

It is quite well-known that fermionic operators provide a representation for the Lie algebra $\mathfrak{so}(N)$ \cite{zhang1990hsuan}. $\mathfrak{so}(2N)$ is defined as the Lie algebra of antisymmetric $2N\times2N$ matrices. Consider the $N(2N-1)$ operators $c_i^{\dagger}c_j-\frac{1}{2}\delta_{ij}$, $i,j=1,2,...,N$ and $c_ic_j(c_i^{\dagger}c_j^{\dagger})$, $i\neq j=1,2,...,N$. In the standard Cartan basis, the commutation relations are 
\begin{align}\label{SO2N commutation relations cartan}
[H_i,c_jc_k]&=-(\delta_{ij}+\delta_{ik})c_jc_k,\\
[H_i,c_j^{\dagger}c_k^{\dagger}]&=(\delta_{ij}+\delta_{ik})c_j^{\dagger}c_k^{\dagger},\\
[H_i,c_j^{\dagger}c_k]&=(\delta_{ij}-\delta_{ik})(c_j^{\dagger}c_k-\frac{1}{2}\delta_{jk}),
\end{align}
where $H_i=c_i^{\dagger}c_i-\frac{1}{2}$ is the Cartan subalgebra. It is easy to see that the corresponding root space is $D_{N}$ which is that of the $\mathfrak{so}(2N)$ algebra.

Similarly, $\mathfrak{so}(2N+1)$ is defined as the Lie algebra of antisymmetric $(2N+1)\times(2N+1)$ matrices. Consider the $N(2N+1)$ operators $c_i^{\dagger}c_j-\frac{1}{2}\delta_{ij}$, $i,j=1,2,...,N$ and $c_ic_j(c_i^{\dagger}c_j^{\dagger})$, $i\neq j=1,2,...,N$, plus the operators $c_i$ and $c_i^{\dagger}$, $i=1,2,...,N$. In other words, we have the generators of the $\mathfrak{so}(2N)$ algebra plus the creation and annihilation operators. In the standard Cartan basis, the commutation relations are the same as those of the $\mathfrak{so}(2N)$ algebra plus
\begin{align}\label{SO2N plus one commutation relations cartan}
[H_i,c_j]&=-\delta_{ij}c_i,\\
[H_i,c_j^{\dagger}]&=\delta_{ij}c_i^{\dagger},
\end{align}
where $H_i=c_i^{\dagger}c_i-\frac{1}{2}$ is the Cartan subalgebra. It is easy to see that the corresponding root space is $B_{N}$ which is that of the $\mathfrak{so}(2N+1)$ algebra. Having this correspondence, one can straightforwardly use the Balian-Brezin formula and its generalization to the cases with the linear part to generate all sorts of BCH (Zassenhaus) formulas for the $\mathfrak{so}(2N)$ and $\mathfrak{so}(2N+1)$ algebra.

\subsection{A non-linear canonical transformation}\label{subsec: non-linear canonical transformation}
In this section, we show how the creation and annihilation operators change under the non-unitary transformation $\mathcal{F}_{(\mathbf{M},\mathbf{u},\mathbf{v})}$, i.e. $\mathcal{F}_{(\mathbf{M},\mathbf{u},\mathbf{v})}^{-1} \begin{pmatrix} \mathbf{c} \\ \mathbf{c}^{\dagger}  \end{pmatrix}\mathcal{F}_{(\mathbf{M},\mathbf{u},\mathbf{v})}$. Using the Colpa trick it is easy to see that:
\begin{equation}
\begin{split}
\mathcal{F}_{\mathbf{M}^{\prime}}^{-1} \begin{pmatrix} c_0 \\ \mathbf{c} \\ c_0^{\dagger} \\ \mathbf{c}^{\dagger}  \end{pmatrix}\mathcal{F}_{\mathbf{M}^{\prime}}=\mathbf{T}^{\prime} \begin{pmatrix} c_0 \\ \mathbf{c} \\ c_0^{\dagger} \\ \mathbf{c}^{\dagger}  \end{pmatrix},
\end{split}
\end{equation}
where $\mathbf{T}'$ is the equation \eqref{eq: T prime matrix}. After lengthy but straightforward calculation one can show that:
\begin{equation}\label{eq: generalized BCH}
\begin{split}
\mathcal{F}_{(\mathbf{M},\mathbf{u},\mathbf{v})}^{-1} \begin{pmatrix} \mathbf{c} \\ \mathbf{c}^{\dagger}  \end{pmatrix}\mathcal{F}_{(\mathbf{M},\mathbf{u},\mathbf{v})}=\mathbf{T}_p \begin{pmatrix} \mathbf{c} \\ \mathbf{c}^{\dagger}  \end{pmatrix} + \begin{pmatrix}  \frac{1}{2} \begin{pmatrix} \mathbf{c}^{\dagger} & \mathbf{c}\\ \end{pmatrix} \Vec{\mathbf{B}} \begin{pmatrix} \mathbf{c} \\ \mathbf{c}^{\dagger}  \end{pmatrix} \\  \frac{1}{2} \begin{pmatrix} \mathbf{c}^{\dagger} & \mathbf{c}\\ \end{pmatrix} \vec{\bar{\mathbf{B}}} \begin{pmatrix} \mathbf{c} \\ \mathbf{c}^{\dagger}  \end{pmatrix} \end{pmatrix}+(1+2t_{11}) \begin{pmatrix} \vec{\mathbf{t}_3} \\ \vec{\mathbf{t}_4}  \end{pmatrix},
\end{split}
\end{equation}
where:
\begin{equation}
\mathbf{T}_p=\left[(1+2t_{11}) \begin{pmatrix} \mathbf{T}_{11} & \mathbf{T}_{12} \\ \mathbf{T}_{21} & \mathbf{T}_{22} \end{pmatrix} -2  \begin{pmatrix} \vec{\mathbf{t}_3} \\ \vec{\mathbf{t}_4}  \end{pmatrix}  \begin{pmatrix} \vec{\mathbf{t}_1}^T & \vec{\mathbf{t}_2}^T  \end{pmatrix} \right], \ \ \ \ \ \begin{cases}
    \vec{\mathbf{B}}^T= \left(\mathbf{B}^1,\mathbf{B}^2,\cdots, \mathbf{B}^L\right)\\
    \vec{\bar{\mathbf{B}}}^T= \left(\bar{\mathbf{B}}^1,\bar{\mathbf{B}}^2,\cdots, \bar{\mathbf{B}}^L\right)
\end{cases},
\end{equation}
and
\begin{equation}
\begin{split}
&\mathbf{B}^{\mu}= \begin{pmatrix} \mathbf{B}_{11}^{\mu} & \mathbf{B}_{12}^{\mu} \\ \mathbf{B}_{21}^{\mu} & \mathbf{B}_{22}^{\mu} \end{pmatrix}, \ \ \ \ \ \ \ \ \bar{\mathbf{B}}^{\mu}= \begin{pmatrix} \bar{\mathbf{B}}_{11}^{\mu} & \bar{\mathbf{B}}_{12}^{\mu} \\ \bar{\mathbf{B}}_{21}^{\mu} & \bar{\mathbf{B}}_{22}^{\mu} \end{pmatrix},
\end{split}
\end{equation}
where
\begin{equation}
\begin{split}
& \begin{pmatrix} (\mathbf{B}_{11}^{\mu})_{\alpha \gamma} & (\mathbf{B}_{12}^{\mu})_{\alpha \gamma} \\ (\mathbf{B}_{21}^{\mu})_{\alpha \gamma} & (\mathbf{B}_{22}^{\mu})_{\alpha \gamma} \end{pmatrix} = -4 \begin{pmatrix} (\vec{\mathbf{t}_2}^T)_{\alpha}(\mathbf{T}_{11})_{\gamma \mu} & (\vec{\mathbf{t}_2}^T)_{\alpha}(\mathbf{T}_{12})_{\gamma \mu} \\ (\vec{\mathbf{t}_1}^T)_{\alpha}(\mathbf{T}_{11})_{\gamma \mu} & (\vec{\mathbf{t}_1}^T)_{\alpha}(\mathbf{T}_{12})_{\gamma \mu} \end{pmatrix} , \\ &\begin{pmatrix} (\bar{\mathbf{B}}_{11}^{\mu})_{\alpha \beta} & (\bar{\mathbf{B}}_{12}^{\mu})_{\alpha \beta} \\ (\bar{\mathbf{B}}_{21}^{\mu})_{\alpha \beta} & (\bar{\mathbf{B}}_{22}^{\mu})_{\alpha \beta} \end{pmatrix}=4 \begin{pmatrix} (\vec{\mathbf{t}_1}^T)_{\beta}(\mathbf{T}_{22})_{ \mu \alpha} & (\vec{\mathbf{t}_2}^T)_{\beta}(\mathbf{T}_{22})_{\mu \alpha} \\ (\vec{\mathbf{t}_1}^T)_{\beta}(\mathbf{T}_{21})_{\mu \alpha} & (\vec{\mathbf{t}_2}^T)_{\beta}(\mathbf{T}_{21})_{\mu \alpha} \end{pmatrix}.
\end{split}
\end{equation}
After some algebra, one can also show that this transformation is a canonical transformation. It is an example of non-linear canonical transformation.

\subsection{Generalized overlap formula}\label{subsec: Generalized overlap}

In this section, we generalize the results of the overlap of the previous sections to the following state: 
\begin{equation}
\ket{(\mathbf{M}_1,\mathbf{u}_1,\mathbf{v}_1)(\mathbf{I})} =\mathcal{F}_{(\mathbf{M}_1,\mathbf{u}_1,\mathbf{v}_1)} \ket{\mathbf{I}}.
\end{equation}
In other words, we would like to calculate the following overlap:
\begin{equation}\label{eq: overlap with linear}
\bracket{(\mathbf{M}_2,\mathbf{u}_2,\mathbf{v}_2)(\mathbf{J})}{(\mathbf{M}_1,\mathbf{u}_1,\mathbf{v}_1)(\mathbf{I})}=\bra{\mathbf{J}} e^{ \frac{1}{2} \begin{pmatrix} \mathbf{c}^{\dagger} & \mathbf{c}\\ \end{pmatrix} \mathbf{M} \begin{pmatrix} \mathbf{c} \\ \mathbf{c}^{\dagger}  \end{pmatrix}+\mathbf{u}^{\dagger}\mathbf{c}^{\dagger}+\mathbf{v}^T\mathbf{c}} \ket{\mathbf{I}},
\end{equation}
where we have $e^{\mathbf{M}_2^{\prime \dagger}}e^{\mathbf{M}_1^{\prime}}=e^{\mathbf{M}^{\prime}}$, with prime over the matrices indicating the enlarged matrices after adding the ancillary site. To calculate equation \eqref{eq: overlap with linear}, we may first try to decompose the exponential in a normal ordered form and then write the Berezin-Grassmann representation and then try to do the integral. Although this procedure works perfectly and one can perform the integrals explicitly when the $\mathbf{I}$ and $\mathbf{J}$ configurations have the same parity; when the two states have opposite parities the integral does not have well-known form due to the singularity of the matrix $\mathcal{A}_{\mathbf{I}_0\mathbf{J}_0}$. However, there is another elegant method which is based on the fact that in the extended Hilbert space $\mathcal{H}^{\prime}$ with dimension $2^{L+1}$ with ancillary site one can do the projection to the subspace $\mathcal{H}$ with dimension $2^L$, using the following substitution \cite{colpa1979diagonalisation}:
\begin{equation}
\ket{\mathbf{I}}\rightarrow \frac{1}{\sqrt{2}}\left(\ket{0,\mathbf{I}}+\ket{1,\mathbf{I}}\right).
\end{equation}
Using the above the overlap will be:
\begin{equation}
\begin{split}    
\bracket{(\mathbf{M}_2,\mathbf{u}_2,\mathbf{v}_2)(\mathbf{J})}{(\mathbf{M}_1,\mathbf{u}_1,\mathbf{v}_1)(\mathbf{I})}=&\frac{1}{2}\Bigg(\bra{\mathbf{J},0} e^{ \frac{1}{2} \begin{pmatrix} \mathbf{c}^{\dagger} & \mathbf{c}\\ \end{pmatrix} \mathbf{M}^{\prime}\begin{pmatrix} \mathbf{c} \\ \mathbf{c}^{\dagger}  \end{pmatrix}} \ket{0,\mathbf{I}} 
 + \bra{\mathbf{J},0} e^{ \frac{1}{2} \begin{pmatrix} \mathbf{c}^{\dagger} & \mathbf{c}\\ \end{pmatrix} \mathbf{M}^{\prime}\begin{pmatrix} \mathbf{c} \\ \mathbf{c}^{\dagger}  \end{pmatrix}} \ket{1,\mathbf{I}} \\
 &+ \bra{\mathbf{J},1} e^{ \frac{1}{2} \begin{pmatrix} \mathbf{c}^{\dagger} & \mathbf{c}\\ \end{pmatrix} \mathbf{M}^{\prime}\begin{pmatrix} \mathbf{c} \\ \mathbf{c}^{\dagger}  \end{pmatrix}} \ket{0,\mathbf{I}}+ \bra{\mathbf{J},1} e^{ \frac{1}{2} \begin{pmatrix} \mathbf{c}^{\dagger} & \mathbf{c}\\ \end{pmatrix} \mathbf{M}^{\prime}\begin{pmatrix} \mathbf{c} \\ \mathbf{c}^{\dagger}  \end{pmatrix}} \ket{1,\mathbf{I}}\Bigg).
\end{split}
\end{equation}
Two of the above terms are always zero because of the parity number symmetry. The other two are always equal so we finally have:
\begin{equation}\label{eq: generalized overlap}
\begin{split}    
\bracket{(\mathbf{M}_2,\mathbf{u}_2,\mathbf{v}_2)(\mathbf{J})}{(\mathbf{M}_1,\mathbf{u}_1,\mathbf{v}_1)(\mathbf{I})}= \begin{cases}
    \bra{\mathbf{J},0} e^{ \frac{1}{2} \begin{pmatrix} \mathbf{c}^{\dagger} & \mathbf{c}\\ \end{pmatrix} \mathbf{M}^{\prime}\begin{pmatrix} \mathbf{c} \\ \mathbf{c}^{\dagger}  \end{pmatrix}} \ket{0,\mathbf{I}}, \ \ \ \ \ \ \ \ \ (-1)^{|\mathbf{I}|}=(-1)^{|\mathbf{J}|}, \\
    \bra{\mathbf{J},0} e^{ \frac{1}{2} \begin{pmatrix} \mathbf{c}^{\dagger} & \mathbf{c}\\ \end{pmatrix} \mathbf{M}^{\prime}\begin{pmatrix} \mathbf{c} \\ \mathbf{c}^{\dagger}  \end{pmatrix}} \ket{1,\mathbf{I}}, \ \ \ \ \ \ \ \ \ (-1)^{|\mathbf{I}|}=(-1)^{|\mathbf{J}|+1}. 
\end{cases}
\end{split}
\end{equation}
Each of the above terms can be calculated using the equation \eqref{eq: reduced grassman integral}. 

For example, consider the following state:
\begin{equation}    
\ket{\psi}=e^{\frac{1}{2}\mathbf{c}^{\dagger}\mathbf{R}\mathbf{c}^{\dagger}+\vec{\mathbf{u}}^{\dagger}\mathbf{c}^{\dagger}} \ket{\vec{\mathbf{0}}}
\end{equation}
Using equation \eqref{eq: generalized overlap} and after some simplifications, we have:
\begin{equation}    
\bracket{\mathbf{J}}{\psi}=\begin{cases}
    \text{pf}\mathbf{R}_{\mathbf{J}_0\mathbf{J}_0}, \ \ \ \ \ \ \ \ \ (-1)^{|\mathbf{J}|}=1,\\
    \text{pf} \mathcal{R}_{\mathbf{J}_0\mathbf{J}_0}, \ \ \ \ \ \ \ \ \ (-1)^{|\mathbf{J}|}=-1,
\end{cases} \ \ \ \ \ \ \ \mathcal{R}=\begin{pmatrix} \mathbf{R} & \vec{\mathbf{u}}^{\ast} \\ -\vec{\mathbf{u}}^{\dagger}&0,  \end{pmatrix},
\end{equation}
where $\mathbf{R}_{\mathbf{J}_0\mathbf{J}_0}$ is the matrix $\mathbf{R}$ in which we remove the rows and columns $\mathbf{J}_0$ and $\mathcal{R}_{\mathbf{J}_0\mathbf{J}_0}$ is the matrix $\mathcal{R}$ in which we remove the rows associated to $\mathbf{J}_0$ and we do the same for columns. The normalization can be also calculated using the equation \eqref{eq: generalized overlap} as:
\begin{equation}    
\bracket{\psi}{\psi}^{2}=\det \begin{pmatrix} \vec{\mathbf{u}}^T.\vec{\mathbf{u}}^{\ast}+1 & -\vec{\mathbf{u}}^T \mathbf{R}-(\vec{\mathbf{u}}^T.\vec{\mathbf{u}}^{\ast}+1)\vec{\mathbf{u}}^{\dagger} \\ -\mathbf{R}^{\dagger}\vec{\mathbf{u}}^{\ast}-\vec{\mathbf{u}}(\vec{\mathbf{u}}^T.\vec{\mathbf{u}}^{\ast}+1)&\mathbf{R}^{\dagger}\mathbf{R}+2\vec{\mathbf{u}}\vec{\mathbf{u}}^{\dagger}+\vec{\mathbf{u}}\vec{\mathbf{u}}^T\mathbf{R}+\mathbf{R}^{\dagger}\vec{\mathbf{u}}^{\ast}\vec{\mathbf{u}}^{\dagger}+\vec{\mathbf{u}}\vec{\mathbf{u}}^T\vec{\mathbf{u}}^{\ast}\vec{\mathbf{u}}^{\dagger}+\mathbf{I} \end{pmatrix}.
\end{equation}
After further simplification we have:
\begin{equation}    
\bracket{\psi}{\psi}^{2}=\frac{1}{(\vec{\mathbf{u}}^T.\vec{\mathbf{u}}^{\ast}+1)^{L-1}}\det\left[(\mathbf{I}+\vec{\mathbf{u}}^T.\vec{\mathbf{u}}^{\ast})(\mathbf{I}+\vec{\mathbf{u}}\vec{\mathbf{u}}^{\dagger}) + \mathbf{R}^{\dagger}(\mathbf{I}+\vec{\mathbf{u}}^T.\vec{\mathbf{u}}^{\ast}\mathbf{I}-\vec{\mathbf{u}}^{\ast}\vec{\mathbf{u}}^T)\mathbf{R}\right] .
\end{equation}
Note that for $\vec{\mathbf{u}}^{\dagger}=0$ we get the expected result $\bracket{\psi}{\psi}^{2}=\det\left(\mathbf{I+\mathbf{R}^{\dagger}\mathbf{R}}\right)$.
\section{Generalized Wick's formula}\label{sec: Generalized Wick}

Consider the following expectation value:
\begin{equation}    
\langle \langle A \rangle \rangle :=\bra{(\mathbf{M}_2,\mathbf{u}_2,\mathbf{v}_2)(\mathbf{J})}A\ket{(\mathbf{M}_1,\mathbf{u}_1,\mathbf{v}_1)(\mathbf{I})},
\end{equation}
where $A$ is an arbitrary product of creation and annihilation operators. We are interested to show that all the expectation values can be written as the combination of the products of one and two-point functions. This can be done by using the Colpa trick and writing: 
\begin{equation}    
\langle \langle A \rangle \rangle =\frac{1}{2}\left(\bra{\mathbf{J},0} + \bra{\mathbf{J},1}\right)e^{ \frac{1}{2} \begin{pmatrix} \mathbf{c}^{\dagger} & \mathbf{c}\\ \end{pmatrix} \mathbf{M}_2^{\prime \dagger}\begin{pmatrix} \mathbf{c} \\ \mathbf{c}^{\dagger}  \end{pmatrix}} 
 A^{\prime} e^{ \frac{1}{2} \begin{pmatrix} \mathbf{c}^{\dagger} & \mathbf{c}\\ \end{pmatrix} \mathbf{M}_1^{\prime}\begin{pmatrix} \mathbf{c} \\ \mathbf{c}^{\dagger}  \end{pmatrix}} 
 \left(\ket{\mathbf{I},0}+ \ket{\mathbf{I},1}\right),
\end{equation}
where $A'$ is the operator $A$ after the substitution \eqref{eq: ancillary site} . As the case of the overlap in subsection \ref{subsec: Generalized overlap}, depending on the  parities of the states $\ket{\mathbf{I}}$ and $\bra{\mathbf{J}}$ we have:
\begin{equation}\label{generalized Wick}
\begin{split}    
\langle \langle A \rangle \rangle= \begin{cases}
    \bra{\mathbf{J},0} e^{ \frac{1}{2} \begin{pmatrix} \mathbf{c}^{\dagger} & \mathbf{c}\\ \end{pmatrix} \mathbf{M}_2^{\prime \dagger}\begin{pmatrix} \mathbf{c} \\ \mathbf{c}^{\dagger}  \end{pmatrix}}  A^{\prime}  e^{ \frac{1}{2} \begin{pmatrix} \mathbf{c}^{\dagger} & \mathbf{c}\\ \end{pmatrix} \mathbf{M}_1^{\prime}\begin{pmatrix} \mathbf{c} \\ \mathbf{c}^{\dagger}  \end{pmatrix}} 
 \ket{0,\mathbf{I}}, \ \ \ \ \ \ \ \ \ (-1)^{|\mathbf{I}|}=(-1)^{|\mathbf{J}|}, \\
    \bra{\mathbf{J},0} e^{ \frac{1}{2} \begin{pmatrix} \mathbf{c}^{\dagger} & \mathbf{c}\\ \end{pmatrix} \mathbf{M}_2^{\prime \dagger}\begin{pmatrix} \mathbf{c} \\ \mathbf{c}^{\dagger}  \end{pmatrix}}  A^{\prime} e^{ \frac{1}{2} \begin{pmatrix} \mathbf{c}^{\dagger} & \mathbf{c}\\ \end{pmatrix} \mathbf{M}_1^{\prime}\begin{pmatrix} \mathbf{c} \\ \mathbf{c}^{\dagger}  \end{pmatrix}} 
 \ket{1,\mathbf{I}}, \ \ \ \ \ \ \ \ \ (-1)^{|\mathbf{I}|}=(-1)^{|\mathbf{J}|+1} .
\end{cases}
\end{split}
\end{equation}
One can now straightforwardly use the Balian-Brezin version of the Wick's theorem to show that the following more general version of Wick's theorem is valid. The expectation value $\langle \langle A \rangle \rangle$ is equal to the sum of all possible completely contracted products with appropriate signs in which appear contractions involving only a pair of operators plus the remaining contraction of a single operator. A few examples are:
\begin{equation}\label{eq: example of wick theorem}
\begin{split}    
&\langle \langle  \phi_i \phi_j \phi_k \rangle  \rangle= \frac{1}{\bracket{(\mathbf{M}_2,\mathbf{u}_2,\mathbf{v}_2)(\mathbf{J})}{(\mathbf{M}_1,\mathbf{u}_1,\mathbf{v}_1)(\mathbf{I})}} \Bigg[\langle\langle \phi_i \phi_j \rangle\rangle \langle\langle \phi_k \rangle\rangle- \langle\langle \phi_i \phi_k \rangle\rangle \langle\langle\phi_j \rangle\rangle + \langle  \langle \phi_j \phi_k \rangle\rangle \langle\langle \phi_i \rangle\rangle \Bigg],\\
&\langle \langle \phi_i \phi_j \phi_k \phi_l\rangle \rangle= \frac{1}{\bracket{(\mathbf{M}_2,\mathbf{u}_2,\mathbf{v}_2)(\mathbf{J})}{(\mathbf{M}_1,\mathbf{u}_1,\mathbf{v}_1)(\mathbf{I})}} \Bigg[ \langle\langle \phi_i \phi_j \rangle\rangle \langle\langle \phi_k \phi_l \rangle\rangle -\langle\langle \phi_i \phi_k \rangle\rangle \langle\langle \phi_j \phi_l \rangle \rangle+ \langle\langle \phi_i \phi_l \rangle\rangle \langle\langle \phi_j \phi_k \rangle\rangle\Bigg],\\
&\langle \langle \phi_i \phi_j \phi_k \phi_l \phi_t \rangle\rangle= \frac{1}{\big[\bracket{(\mathbf{M}_2,\mathbf{u}_2,\mathbf{v}_2)(\mathbf{J})}{(\mathbf{M}_1,\mathbf{u}_1,\mathbf{v}_1)(\mathbf{I})}\big]^2} \Bigg[ \langle \langle \phi_i \phi_j \rangle\rangle \langle\langle \phi_k \phi_l \rangle\rangle \langle\langle \phi_t \rangle\rangle -\langle\langle \phi_i \phi_k \rangle\rangle \langle\langle \phi_j \phi_l \rangle\rangle \langle\langle \phi_t \rangle\rangle \\& + \langle \langle \phi_i \phi_l \rangle\rangle \langle\langle \phi_j \phi_k \rangle\rangle \langle\langle \phi_t \rangle\rangle -\langle\langle \phi_i \phi_j \rangle\rangle \langle\langle \phi_k \phi_t \rangle\rangle \langle\langle \phi_l \rangle\rangle + \langle\langle \phi_i \phi_k \rangle\rangle \langle\langle \phi_j \phi_t \rangle\rangle \langle\langle \phi_l \rangle\rangle - \langle\langle \phi_i \phi_t \rangle\rangle \langle\langle \phi_j \phi_k \rangle\rangle \langle\langle \phi_l \rangle\rangle  \\ &+\langle\langle \phi_i \phi_j \rangle\rangle \langle\langle \phi_l \phi_t \rangle\rangle \langle\langle \phi_k \rangle\rangle -\langle\langle \phi_i \phi_l \rangle\rangle \langle\langle \phi_j \phi_t \rangle\rangle \langle\langle \phi_k \rangle\rangle + \langle\langle \phi_i \phi_t \rangle\rangle \langle\langle \phi_j \phi_t \rangle\rangle \langle\langle \phi_k \rangle\rangle -\langle\langle \phi_i \phi_k \rangle\rangle \langle\langle \phi_l \phi_t \rangle\rangle \langle\langle \phi_j \rangle\rangle \\& + \langle\langle \phi_i \phi_l \rangle\rangle \langle\langle \phi_k \phi_t \rangle\rangle \langle\langle \phi_j \rangle\rangle - \langle\langle \phi_i \phi_t \rangle\rangle \langle\langle \phi_k \phi_l \rangle\rangle \langle\langle \phi_j \rangle\rangle +\langle\langle \phi_j \phi_k \rangle\rangle \langle\langle \phi_l \phi_t \rangle\rangle \langle\langle \phi_i \rangle\rangle -\langle\langle \phi_j \phi_j \rangle\rangle \langle\langle \phi_k \phi_t \rangle\rangle \langle\langle \phi_i \rangle\rangle\\& + \langle\langle \phi_j \phi_t \rangle\rangle \langle\langle \phi_k \phi_l \rangle\rangle \langle\langle \phi_i \rangle\rangle\Bigg].
\end{split}
\end{equation}
Note that the one and two-point functions can be derived using the equations of subsection \ref{subsec: Correlation functions}.

\section{Conclusions}\label{sec: Conclusions}

In this paper, we introduced the fermionic Gaussian operators with a linear part and demonstrated their decomposition into five exponentials, akin to the treatment of operators without the linear term in the work of Balian and Brezin \cite{balian1969nonunitary}. Correspondingly, we defined Gaussian states associated with these operators and explicitly calculated the overlap formulas, expressed as the Pfaffian of a block matrix. These formulas not only yield the matrix elements of the fermionic Gaussian operators with a linear part in the configuration basis but also offer a solution for cases where the matrix $\mathbf{T}_{22}$ lacks an inverse. Additionally, we established a generalization of the Balian-Brezin version of Wick's theorem for these more general states.

The derived formulas hold promise for applications in various domains, such as the study of quantum spin chains featuring longitudinal magnetic fields at the boundary. Moreover, they may serve as effective approximations for the ground states of quantum chains lacking parity number symmetry. Furthermore, these results can find utility in the analysis of dissipative systems described by the Lindblad equation.
\newline
\newline

{\textit{Acknowledgment}}:
The work done by ASM was funded by the CAPES. MAR thanks CNPq and FAPERJ (grant number 26/210.040/2020) for partial support. The work done by AJ was funded by the Royal Commission for the Exhibition of 1851.

 \appendix

\section{An example}\label{sec: an example}

To illustrate the BB decomposition and how to handle the possible singularities of submatrices of the $\mathbf{T}$ matrix, in this section, we provide an example which can be treated entirely in analytical form.
Consider the following operator:

\begin{equation}\label{eq: F example}
    \mathcal{F}=e^{(-\text{M}_{12}c_1^{\dagger}c_2+\text{M}_{12}c_1^{\dagger}c_3^{\dagger} + \text{M}_{12}c_3c_2^{\dagger}+\text{M}_{12}c_1c_3)},
\end{equation} 
where the $\mathbf{M}$ matrix and corresponding $\mathbf{T}$ matrix are as follows:
 \begin{equation}
\begin{split}
& \mathbf{M}=\left(
\begin{array}{cccccc}
 0 & -\text{M}_{12} & 0 & 0 & 0 & \text{M}_{12} \\
 0 & 0 & -\text{M}_{12} & 0 & 0 & 0 \\
 0 & 0 & 0 & -\text{M}_{12} & 0 & 0 \\
 0 & 0 & \text{M}_{12} & 0 & 0 & 0 \\
 0 & 0 & 0 & \text{M}_{12} & 0 & 0 \\
 -\text{M}_{12} & 0 & 0 & 0 & \text{M}_{12} & 0 \\
\end{array}
\right),\\
& \mathbf{T}=\left(
\begin{array}{cccccc}
 \cos (\text{M}_{12}) & -\sin (\text{M}_{12}) & 1-\cos (\text{M}_{12}) & 0 & 1-\cos (\text{M}_{12}) & \sin (\text{M}_{12}) \\
 0 & 1 & -\sin (\text{M}_{12}) & 1-\cos (\text{M}_{12}) & 0 & 0 \\
 0 & 0 & \cos (\text{M}_{12}) & -\sin (\text{M}_{12}) & 0 & 0 \\
 0 & 0 & \sin (\text{M}_{12}) & \cos (\text{M}_{12}) & 0 & 0 \\
 0 & 0 & 1-\cos (\text{M}_{12}) & \sin (\text{M}_{12}) & 1 & 0 \\
 -\sin (\text{M}_{12}) & 1-\cos (\text{M}_{12}) & 0 & 1-\cos (\text{M}_{12}) & \sin (\text{M}_{12}) & \cos (\text{M}_{12}) \\
\end{array}
\right).
\end{split}
\end{equation}

Note that the $\mathbf{J}.\mathbf{M}$ is antisymmetric.
When the determinant of sub-matrices $\mathbf{T}_{11}$ and $\mathbf{T}_{22}$ are non-zero, then one can  apply BB decomposition to equation \eqref{eq: F example}. The possible decompositions are written in the table \ref{eq: table3}.
\begin{table}[h]
\centering
\caption{\footnotesize Balian Brezin Decomposition for invertible $\mathbf{T}_{22}$ and $\mathbf{T}_{11}$ submatrices for $L=3$ \label{eq: table3}.}
\renewcommand{\arraystretch}{1.9}
\begin{tabular}{|c|c|}
    \hline
    {\footnotesize Invertibility} &  \begin{tabular}{@{}c@{}} {\footnotesize BB Decomposotion} \\ $e^{(-\text{M}_{12}c_1^{\dagger}c_2+\text{M}_{12}c_1^{\dagger}c_3^{\dagger} + \text{M}_{12}c_3c_2^{\dagger}+\text{M}_{12}c_1c_3)}$ \end{tabular} \\
    \hline
    $\mathbf{T}_{22}^{-1}$ exist & \tiny \begin{tabular}{@{}c@{}}  $e^{(1-\sec{(\text{M}_{12})}c_1^{\dagger}c_2^{\dagger}+\tan{\text{M}_{12}}c_1^{\dagger}c_3^{\dagger})}$ \\  $e^{(-\log{(\cos(\text{M}_{12}))}c_1^{\dagger}c_1 +\cot \left(\frac{\text{M}_{12}}{2}\right) \log (\cos (\text{M}_{12})) c_1^{\dagger}c_2 +(\cot ^2\left(\frac{\text{M}_{12}}{2}\right) \log (\cos (\text{M}_{12}))+2)c_1^{\dagger}c_3+ \cot \left(\frac{\text{M}_{12}}{2}\right) \log (\cos (\text{M}_{12})) c_2^{\dagger}c_3-\log(\cos(\text{M}_{12})) c_3^{\dagger} c_2)} $  \\  $ e^{(1-\sec{(\text{M}_{12})}c_2c_3+\tan{\text{M}_{12}}c_1c_3)} e^{\log{(\cos(\text{M}_{12}))}}$\end{tabular} \\
    \hline
    $\mathbf{T}_{11}^{-1}$ exist &\tiny \begin{tabular}{@{}c@{}} $e^{(-1+\sec{(\text{M}_{12})}c_2c_3+\tan{\text{M}_{12}}c_1c_3)}$ \\ $ e^{(\log{(\cos(\text{M}_{12}))}c_1^{\dagger}c_1 +\cot \left(\frac{\text{M}_{12}}{2}\right) \log (\cos (\text{M}_{12})) c_1^{\dagger}c_2 -(\cot ^2\left(\frac{\text{M}_{12}}{2}\right) \log (\cos (\text{M}_{12}))+2)c_1^{\dagger}c_3+ \cot \left(\frac{\text{M}_{12}}{2}\right) \log (\cos (\text{M}_{12})) c_2^{\dagger}c_3+\log(\cos(\text{M}_{12})) c_3^{\dagger} c_2)}$  \\ $e^{(-1+\sec{(\text{M}_{12})}c_1^{\dagger}c_2^{\dagger}+\tan{\text{M}_{12}}c_1^{\dagger}c_3^{\dagger})} e^{-\log{(\cos(\text{M}_{12}))}}$\end{tabular}
  \\
    \hline
\end{tabular}
\end{table}\\
Choosing $\text{M}_{12}=\frac{(2k+1)\pi}{2}$, the determinant of $\mathbf{T}_{11}$ and $\mathbf{T}_{22}$ will be zero, consequently they are not invertible and one can not use BBD. We show now how to use  the permutation matrix technique to change the submatrices and get a non-zero value for the determinant of $\mathbf{T}_{11}$ and $\mathbf{T}_{22}$ matrices. This permutation matrix is   $c_j\rightarrow \Tilde{c}_j^{\dagger}$ and $c_j^{\dagger}\rightarrow \Tilde{c}_j$ performed on one or more sites. One can see all the possible transformations and corresponding BB decomposition in the table \ref{table2}.\\
 \begin{table}[h]
\centering
\caption{\footnotesize Balian Brezin Decomposition for invertible $\mathbf{T}_{22}$ and $\mathbf{T}_{11}$ submatrices after applying the canonical permutation transformation for $L=3$.\label{table2}}
\renewcommand{\arraystretch}{1.9}
\begin{tabular}{|c|c|c|}
    \hline
  {\footnotesize CP Transformation} & {\footnotesize Invertibility} & \begin{tabular}{@{}c@{}} {\footnotesize BB Decomposition }\\ $e^{\frac{\pi}{2}(-c_1^{\dagger}c_2+c_1^{\dagger}c_3^{\dagger}+c_3c_2^{\dagger} + c_1c_3)}$ \end{tabular}\\
    \hline
   \begin{tabular}{@{}c@{}} $c_1\leftarrow\!\rightarrow\tilde{c}_1^\dagger$ \end{tabular} &$\Tilde{\mathbf{T}}_{22}^{-1}$ exist &   $e^{\frac{\pi}{2}(c_1c_3 -  c_2^{\dagger}c_3 - c_3^{\dagger}c_1^{\dagger})}e^{(-c_1^{\dagger}c_2+c_1^{\dagger}c_3+c_2c_3)}$ \\
    \cline{2-3}
     {}& $\Tilde{\mathbf{T}}_{11}^{-1}$ exist &  $e^{(-c_1^{\dagger}c_2+c_3c_1^{\dagger}-c_2c_3)}e^{\frac{\pi}{2}(c_1c_3 -  c_2^{\dagger}c_3 - c_3^{\dagger}c_1^{\dagger})}$
  \\
  \hline
       \begin{tabular}{@{}c@{}}   $c_2\leftarrow\!\rightarrow\tilde{c}_2^\dagger$ \end{tabular} & {\footnotesize $\Tilde{\mathbf{T}}_{22}$ and $\Tilde{\mathbf{T}}_{11}$ are singular} &   {}
  \\
  \hline
     \begin{tabular}{@{}c@{}} $c_3\leftarrow\!\rightarrow\tilde{c}_3^\dagger$ \end{tabular}  &$\Tilde{\mathbf{T}}_{22}^{-1}$ exist &  $e^{(c_1^{\dagger}c_2^{\dagger}+c_1^{\dagger}c_3-c_2^{\dagger}c_3)}e^{\frac{\pi}{2}(-c_3c_1 -  c_1^{\dagger}c_2 + c_1^{\dagger}c_3^{\dagger})}$ \\
    \cline{2-3}
     {}& $\Tilde{\mathbf{T}}_{11}^{-1}$ exist &  $ e^{\frac{\pi}{2}(-c_3c_1 -  c_1^{\dagger}c_2 + c_1^{\dagger}c_3^{\dagger})}e^{(-c_1^{\dagger}c_2^{\dagger}-c_1^{\dagger}c_3-c_2^{\dagger}c_3)}$
  \\
  \hline
     \begin{tabular}{@{}c@{}} $c_1\leftarrow\!\rightarrow\tilde{c}_1^\dagger$ \\ $c_2\leftarrow\!\rightarrow\tilde{c}_2^\dagger$ \end{tabular}  &$\Tilde{\mathbf{T}}_{22}^{-1}$ exist & $ e^{\frac{\pi}{2}(c_2c_1^{\dagger}-c_3^{\dagger}c_1^{\dagger}+c_1c_3)}e^{(-c_1^{\dagger}c_2^{\dagger} -  c_1^{\dagger}c_3 - c_2^{\dagger}c_3)}$ \\
    \cline{2-3}
    {} &$\Tilde{\mathbf{T}}_{11}^{-1}$ exist & $ e^{(c_1^{\dagger}c_2^{\dagger} +  c_1^{\dagger}c_3 - c_2^{\dagger}c_3)} e^{\frac{\pi}{2}(c_2c_1^{\dagger}-c_3^{\dagger}c_1^{\dagger}+c_1c_3)}$ \\
    \hline
     \begin{tabular}{@{}c@{}}$c_1\leftarrow\!\rightarrow\tilde{c}_1^\dagger$ \\ $c_3\leftarrow\!\rightarrow\tilde{c}_3^\dagger$\end{tabular}   &{\footnotesize $\Tilde{\mathbf{T}}_{22}$ and $\Tilde{\mathbf{T}}_{11}$ are singular} & {} \\
    \hline
    \begin{tabular}{@{}c@{}} $c_2\leftarrow\!\rightarrow\tilde{c}_2^\dagger$ \\ $c_3\leftarrow\!\rightarrow\tilde{c}_3^\dagger$ \end{tabular} &$\Tilde{\mathbf{T}}_{22}^{-1}$ exist &  $e^{(-c_1^{\dagger}c_2-c_1^{\dagger}c_3-c_2c_3)}e^{\frac{\pi}{2}(-c_3c_1 +  c_2c_2^{\dagger} + c_1^{\dagger}c_3^{\dagger})}$ \\
    \cline{2-3}
     {} &$\Tilde{\mathbf{T}}_{11}^{-1}$ exist &  $e^{\frac{\pi}{2}(-c_3c_1 +  c_2c_2^{\dagger} + c_1^{\dagger}c_3^{\dagger})} e^{(-c_1^{\dagger}c_2+c_1^{\dagger}c_3+c_2c_3)} $
  \\
    \hline
\end{tabular}
\end{table}
It is worth mentioning that there is no guarantee that after the transformation the diagonal submatrices of $\mathbf{T}$ always have inverses. In some cases, the original matrix might have an invertible submatrix, but it may not be invertible after transformation. This just means that the corresponding decomposition is not possible. Note that one can follow the same procedure and decompose further the exponentials produced after the first step and continue such that each exponential has just one of the terms $\{cc,c^{\dagger}c,c^{\dagger}c^{\dagger}\}$. Some examples are provided in the tables \ref{table1} and \ref{table11}.

\subsection{Matrix elements}\label{subsec: Matrix elements}

In this subsection, we provide the matrix elements of the operator (\ref{eq: F example}). Using the equation (\ref{eq: reduced grassman integral}) we have:
\begin{equation}\label{eq: matrix element}
\mathcal{F}=\left(
\begin{array}{cccccccc}
 \cos (a) & 0 & 0 & 0 & 0 & -\sin (a) &1-\cos (a) & 0 \\
 0 & 1 & -\sin (a) & 0 & 1-\cos (a) & 0 & 0 & -1+\cos (a) \\
 0 & 0 & \cos (a) & 0 & -\sin (a) & 0 & 0 & \sin (a) \\
 \cos (a)-1 & 0 & 0 & 1 & 0 & -\sin (a) & 1-\cos (a) & 0 \\
 0 & 0 & 0 & 0 & 1 & 0 & 0 & 0 \\
 \sin (a) & 0 & 0 & 0 & 0 & \cos (a) & -\sin (a) & 0 \\
 0 & 0 & 0 & 0 & 0 & 0 & 1 & 0 \\
 0 & 0 & -\sin (a) & 0 & 1-\cos (a) & 0 & 0 & \cos (a) \\
\end{array}
\right).
\end{equation}
This should be compared with the matrix that one can get using the Jordan-Wigner transformation ($c_l=\prod_{j<l}\sigma_j^z \sigma_l^-$ and $c_l^{\dagger}=\prod_{j<l}\sigma_j^z \sigma_l^+$), i.e.
\begin{equation}\label{matrix element JW}
\mathcal{F}_{JW}=\left(
\begin{array}{cccccccc}
 \cos (a) & 0 & 0 & 0 & 0 & -\sin (a) & \cos (a)-1 & 0 \\
 0 & 1 & \sin (a) & 0 & 1-\cos (a) & 0 & 0 & 1-\cos (a) \\
 0 & 0 & \cos (a) & 0 & \sin (a) & 0 & 0 & \sin (a) \\
 1-\cos (a) & 0 & 0 & 1 & 0 & \sin (a) & 1-\cos (a) & 0 \\
 0 & 0 & 0 & 0 & 1 & 0 & 0 & 0 \\
 \sin (a) & 0 & 0 & 0 & 0 & \cos (a) & \sin (a) & 0 \\
 0 & 0 & 0 & 0 & 0 & 0 & 1 & 0 \\
 0 & 0 & -\sin (a) & 0 & \cos (a)-1 & 0 & 0 & \cos (a) \\
\end{array}
\right),
\end{equation}
where $\sigma_l^{\pm}=\frac{\sigma_l^x \pm  i\sigma_l^{y}}{2}$ and $\sigma^i (i=x,y,z)$ are the pauli matrices. Note that equation \eqref{eq: matrix element} is correct although for $a=\frac{(2k+1)\pi}{2}$ the BBD does not exist. This shows that in these cases one can assume $\cos{a}\neq 0$ and put it equal to zero just at the end of the calculations. This might be a very useful technique in numerical calculations by just considering a small parameter in calculations.
Also, the sign difference can be traced back to the following relations between the bases in the fermionic and spin versions:
\begin{equation}
\begin{split}
&\ket{000}\equiv\ket{\downarrow \downarrow \downarrow}, \ \ \ \ \ \ \ket{100}\equiv\ket{\uparrow \downarrow \downarrow}, \ \ \ \ \ \ \ket{010}\equiv -\ket{\downarrow \uparrow \downarrow}, \ \ \ \ \ \ \ket{001}\equiv\ket{\downarrow \downarrow \uparrow}, \\
& \ket{110}\equiv -\ket{\uparrow \uparrow \downarrow}, \ \ \ \ \ \ \ket{101}\equiv\ket{\uparrow \downarrow \uparrow}, \ \ \ \ \ \ \ket{011}\equiv -\ket{\downarrow \uparrow \uparrow}, \ \ \ \ \ \ \ket{111}\equiv -\ket{\uparrow \uparrow \uparrow},
\end{split}
\end{equation}
For general $L$ the above correspondence is as follows:
\begin{equation}
\begin{split}
&\ket{\mathbf{0}} \equiv \ket{0\cdots0}\equiv\ket{\downarrow \cdots \downarrow}, \\ 
&\ket{\{i_1,i_2,...,i_n\}} \equiv \ket{\cdots01_{i_1}0\cdots1_{i_2}0..1_{i_3}0\cdots}\equiv\prod_{i\in\{\mathbf{I}\}}c_i^{\dagger} \ket{0\cdots0}
\equiv \prod_{i\in\{\mathbf{I}\}} \prod_{j<i}\sigma_j^z\sigma_i^+\ket{\downarrow \cdots \downarrow}\\
&=(-1)^{\sum_{j=1}^n(i_j-1)}\ket{\cdots\downarrow\uparrow_{i_1}\downarrow\cdots\uparrow_{i_2}\downarrow..\uparrow_{i_3}\downarrow\cdots}.  
\end{split}
\end{equation}
In the above, we assume that we have fermions at the sites $\mathbf{I}=\{i_1,i_2,...,i_n\}$.
    \begin{table}[h]
\centering
\caption{\footnotesize Balian Brezin Decomposition for the first exponential of the first line of table \ref{table2}\label{table1}. }
\renewcommand{\arraystretch}{1.9}
\begin{tabular}{|c|c|c|}
    \hline
   {\footnotesize CP Transformation}& Invertibility &  \begin{tabular}{@{}c@{}} BB Decomposition \\ $e^{\frac{\pi}{2}(c_1c_3 -  c_2^{\dagger}c_3 - c_3^{\dagger}c_1^{\dagger})}$ \end{tabular}\\
    \hline
     \begin{tabular}{@{}c@{}} $c_3\leftarrow\!\rightarrow\tilde{c}_3^\dagger$ \end{tabular} & $\mathbf{T^{\prime}}_{22}^{-1}$ exist &  \begin{tabular}{@{}c@{}}
     $e^{(c_1^{\dagger}c_2^{\dagger}- c_2^{\dagger}c_3)}e^{\frac{\pi}{2}(c_1^{\dagger}c_3^{\dagger} - c_3c_1)} $ \end{tabular} \\
    \cline{2-3}
    {} & $\mathbf{T^{\prime}}_{11}^{-1}$exist &  \begin{tabular}{@{}c@{}}
$ e^{\frac{\pi}{2}(c_1^{\dagger}c_3^{\dagger} - c_3c_1)} e^{(-c_1^{\dagger}c_2^{\dagger}- c_2^{\dagger}c_3)} $ \end{tabular}
  \\
    \hline
\end{tabular}
\end{table}
     \begin{table}[h]
\centering
\caption{\footnotesize Balian Brezin Decomposition for the second exponential of the first line of table \ref{table2}\label{table11}.}
\renewcommand{\arraystretch}{1.9}
\begin{tabular}{|c|c|}
    \hline
    Invertibility &  \begin{tabular}{@{}c@{}} BB Decomposition \\ $e^{(-c_1^{\dagger}c_2 +  c_1^{\dagger}c_3 + c_2c_3)}$ \end{tabular}\\
    \hline
    $\mathbf{T^{\prime}}_{22}^{-1}$ exist &  \begin{tabular}{@{}c@{}}
     $e^{(-c_1^{\dagger}c_2+ c_1^{\dagger}c_3)}e^{c_2c_3} $ \end{tabular} \\
    \hline
    $\mathbf{T^{\prime}}_{11}^{-1}$exist &  \begin{tabular}{@{}c@{}}
     $e^{c_2c_3}e^{(-c_1^{\dagger}c_2+ c_1^{\dagger}c_3)}$ \end{tabular}
  \\
    \hline
\end{tabular}
\end{table}

\section{A Simple example of decomposition of exponential with linear term}\label{sec: Example of decomposition of exponential with linear term}

In this subsection, we present all the possible decompositions of the following exponential:
\begin{equation}\label{eq: example of F}
f_{(a,b,d)}=e^{ac^{\dagger}+bc+\frac{d}{2}\left(2c^{\dagger}c-1\right)}.
\end{equation}
There are $8$ interesting decompositions which are summarized in the table \ref{table5}. The $\alpha$, $\beta$ and $\gamma$ can be derived by solving the equation:
\begin{equation}
e^{\mathbf{M}_1^{\prime}} e^{ \mathbf{M}_2^{\prime}} e^{\mathbf{M}_3^{\prime}}= e^{\mathbf{M}^{\prime}}.
\end{equation}
\begin{table}[h]
\centering
\caption{\footnotesize Possible decompositions of equation \eqref{eq: example of F}\label{table5} }
\renewcommand{\arraystretch}{1.99}
\begin{tabular}{|c|c|c|c|c|}
    \hline
  {\footnotesize Decompositions} & {\footnotesize $\mathbf{M}^{\prime}_1$} & {\footnotesize $\mathbf{M}^{\prime}_2$ } & {\footnotesize $\mathbf{M}^{\prime}_3$ } & {\footnotesize $\mathbf{M}^{\prime}$ }\\
    \hline
    \tiny $e^{\alpha c^{\dagger}}e^{\gamma/2(2c^{\dagger}c-1)}e^{\beta c}$& \tiny $\left(
\begin{array}{cccc}
 0 & 0 & 0 & \alpha \\
 \alpha  & 0 & -\alpha  & 0 \\
 0 & 0 & 0 & -\alpha \\
 0 & 0 & 0 & 0 \\
\end{array}
\right) $& \tiny $\left(
\begin{array}{cccc}
 0 & 0 & 0 & 0 \\
 0 & \gamma & 0 & 0 \\
 0 & 0 & 0 & 0 \\
 0 & 0 & 0 & -\gamma \\
\end{array}
\right)$  & \tiny $\left(
\begin{array}{cccc}
 0 & \beta  & 0 & 0 \\
 0 & 0 & 0 & 0 \\
 0 & -\beta  & 0 & 0 \\
 \beta & 0 & -\beta & 0 \\
\end{array}
\right)$ & {}\\
\cline{1-4}
   \tiny $e^{\gamma/2(2c^{\dagger}c-1)}e^{\alpha c^{\dagger}}e^{\beta c}$& \tiny $\left(
\begin{array}{cccc}
 0 & 0 & 0 & 0 \\
 0 & \gamma & 0 & 0 \\
 0 & 0 & 0 & 0 \\
 0 & 0 & 0 & -\gamma \\
\end{array}
\right)$ & \tiny $\left(
\begin{array}{cccc}
 0 & 0 & 0 & \alpha \\
 \alpha  & 0 & -\alpha  & 0 \\
 0 & 0 & 0 & -\alpha \\
 0 & 0 & 0 & 0 \\
\end{array}
\right) $   & \tiny $\left(
\begin{array}{cccc}
 0 & \beta  & 0 & 0 \\
 0 & 0 & 0 & 0 \\
 0 & -\beta  & 0 & 0 \\
 \beta & 0 & -\beta & 0 \\
\end{array}
\right)$ & {}\\
\cline{1-4}
    \tiny $e^{\beta c}e^{\gamma/2(2c^{\dagger}c-1)}e^{\alpha c^{\dagger}}$& \tiny $\left(
\begin{array}{cccc}
 0 & \beta  & 0 & 0 \\
 0 & 0 & 0 & 0 \\
 0 & -\beta  & 0 & 0 \\
 \beta & 0 & -\beta & 0 \\
\end{array}
\right)$ & \tiny $\left(
\begin{array}{cccc}
 0 & 0 & 0 & 0 \\
 0 & \gamma & 0 & 0 \\
 0 & 0 & 0 & 0 \\
 0 & 0 & 0 & -\gamma \\
\end{array}
\right)$  & \tiny $\left(
\begin{array}{cccc}
 0 & 0 & 0 & \alpha \\
 \alpha  & 0 & -\alpha  & 0 \\
 0 & 0 & 0 & -\alpha \\
 0 & 0 & 0 & 0 \\
\end{array}
\right) $ & \tiny $\left(
\begin{array}{cccc}
 0 & b & 0 & a \\
 a & d & -a & 0 \\
 0 & -b & 0 & -a \\
 b & 0 & -b & -d \\
\end{array}
\right)$\\
    \cline{1-4}
     \tiny $e^{\gamma/2(2c^{\dagger}c-1)}e^{\beta c}e^{\alpha c^{\dagger}}$& \tiny $\left(
\begin{array}{cccc}
 0 & 0 & 0 & 0 \\
 0 & \gamma & 0 & 0 \\
 0 & 0 & 0 & 0 \\
 0 & 0 & 0 & -\gamma \\
\end{array}
\right)$  & \tiny $\left(
\begin{array}{cccc}
 0 & \beta  & 0 & 0 \\
 0 & 0 & 0 & 0 \\
 0 & -\beta  & 0 & 0 \\
 \beta & 0 & -\beta & 0 \\
\end{array}
\right)$   & \tiny $\left(
\begin{array}{cccc}
 0 & 0 & 0 & \alpha \\
 \alpha  & 0 & -\alpha  & 0 \\
 0 & 0 & 0 & -\alpha \\
 0 & 0 & 0 & 0 \\
\end{array}
\right) $ & {}\\
    \cline{1-4}
        \tiny $e^{\alpha c^{\dagger}}e^{\beta c}e^{\gamma/2(2c^{\dagger}c-1)}$& \tiny $\left(
\begin{array}{cccc}
 0 & 0 & 0 & \alpha \\
 \alpha  & 0 & -\alpha  & 0 \\
 0 & 0 & 0 & -\alpha \\
 0 & 0 & 0 & 0 \\
\end{array}
\right) $ & \tiny $\left(
\begin{array}{cccc}
 0 & \beta  & 0 & 0 \\
 0 & 0 & 0 & 0 \\
 0 & -\beta  & 0 & 0 \\
 \beta & 0 & -\beta & 0 \\
\end{array}
\right)$   & \tiny $\left(
\begin{array}{cccc}
 0 & 0 & 0 & 0 \\
 0 & \gamma & 0 & 0 \\
 0 & 0 & 0 & 0 \\
 0 & 0 & 0 & -\gamma \\
\end{array}
\right)$  & {}\\
    \cline{1-4}
       \tiny $e^{\beta c}e^{\alpha c^{\dagger}}e^{\gamma/2(2c^{\dagger}c-1)}$& \tiny $\left(
\begin{array}{cccc}
 0 & \beta  & 0 & 0 \\
 0 & 0 & 0 & 0 \\
 0 & -\beta  & 0 & 0 \\
 \beta & 0 & -\beta & 0 \\
\end{array}
\right)$ & \tiny $\left(
\begin{array}{cccc}
 0 & 0 & 0 & \alpha \\
 \alpha  & 0 & -\alpha  & 0 \\
 0 & 0 & 0 & -\alpha \\
 0 & 0 & 0 & 0 \\
\end{array}
\right) $    & \tiny $\left(
\begin{array}{cccc}
 0 & 0 & 0 & 0 \\
 0 & \gamma & 0 & 0 \\
 0 & 0 & 0 & 0 \\
 0 & 0 & 0 & -\gamma \\
\end{array}
\right)$  & {}\\
\cline{1-4}
       \tiny $e^{\alpha c^{\dagger}+\beta c}e^{\gamma/2(2c^{\dagger}c-1)}$& \tiny $\left(
\begin{array}{cccc}
 0 & \beta  & 0 & \alpha \\
 \alpha & 0 & -\alpha & 0 \\
 0 & -\beta  & 0 & -\alpha \\
 \beta & 0 & -\beta & 0 \\
\end{array}
\right)$ & \tiny $\left(
\begin{array}{cccc}
 0 & 0 & 0 & 0 \\
 0 & \gamma & 0 & 0 \\
 0 & 0 & 0 & 0 \\
 0 & 0 & 0 & -\gamma \\
\end{array}
\right)$     & {} & {}\\
\cline{1-4}
       \tiny $e^{\gamma/2(2c^{\dagger}c-1)} e^{\alpha c^{\dagger}+\beta c}$& \tiny $\left(
\begin{array}{cccc}
 0 & 0 & 0 & 0 \\
 0 & \gamma & 0 & 0 \\
 0 & 0 & 0 & 0 \\
 0 & 0 & 0 & -\gamma \\
\end{array}
\right)$     & \tiny $\left(
\begin{array}{cccc}
 0 & \beta  & 0 & \alpha \\
 \alpha & 0 & -\alpha & 0 \\
 0 & -\beta  & 0 & -\alpha \\
 \beta & 0 & -\beta & 0 \\
\end{array}
\right)$  & {} & {}\\
    \hline
\end{tabular}
\end{table}\\
The results for the first $6$ cases can be derived explicitly, which are summarized in the table \ref{table6}. They should be compared with the decompositions derived for the $\mathfrak{so}(3)$ algebra listed in \cite{gilmore2006lie}.
\begin{table}[h]
\centering
\caption{\footnotesize Relation between $\alpha$, $\beta$ and $\gamma$ with $a$, $b$ and $d$ from the table \ref{table5} \label{table6} }
\renewcommand{\arraystretch}{1.99}
\begin{tabular}{|c|c|c|c|}
    \hline
  {\footnotesize Type} & {\footnotesize $\gamma$} & {\footnotesize $\alpha$ } & {\footnotesize $\beta$ } \\
    \hline
     $I$& \footnotesize		$e^{-\gamma /2}=\cosh \left(\frac{1}{2} \sqrt{4 a b+d^2}\right)-\frac{d \sinh
   \left(\frac{1}{2} \sqrt{4 a b+d^2}\right)}{\sqrt{4 a b+d^2}}$& \footnotesize	 $\alpha= \frac{\frac{2 a \sinh
   \left(\frac{1}{2} \sqrt{4 a b+d^2}\right)}{\sqrt{4 a b+d^2}}}{e^{-\gamma /2}}$ & \footnotesize	$\beta=\frac{\frac{2 b \sinh
   \left(\frac{1}{2} \sqrt{4 a b+d^2}\right)}{\sqrt{4 a b+d^2}}}{e^{-\gamma /2}}$\\
\cline{1-4}
   $II$&\footnotesize	$e^{\gamma /2}=\cosh \left(\frac{1}{2} \sqrt{4 a b+d^2}\right)+\frac{d \sinh
   \left(\frac{1}{2} \sqrt{4 a b+d^2}\right)}{\sqrt{4 a b+d^2}}$ & \footnotesize	$\alpha= \frac{\frac{2 a \sinh
   \left(\frac{1}{2} \sqrt{4 a b+d^2}\right)}{\sqrt{4 a b+d^2}}}{e^{\gamma /2}}$   &\footnotesize	 $\beta=\frac{\frac{2 b \sinh
   \left(\frac{1}{2} \sqrt{4 a b+d^2}\right)}{\sqrt{4 a b+d^2}}}{e^{\gamma /2}}$\\
\cline{1-4}
    $III$&\footnotesize	 $e^{-\gamma /2}=\cosh \left(\frac{1}{2} \sqrt{4 a b+d^2}\right)-\frac{d \sinh
   \left(\frac{1}{2} \sqrt{4 a b+d^2}\right)}{\sqrt{4 a b+d^2}}$ & \footnotesize	$\alpha= \frac{\frac{2 a \sinh
   \left(\frac{1}{2} \sqrt{4 a b+d^2}\right)}{\sqrt{4 a b+d^2}}}{e^{-\gamma /2}}$ &\footnotesize	 $\beta=\frac{2 b \sinh
   \left(\frac{1}{2} \sqrt{4 a b+d^2}\right)}{\sqrt{4 a b+d^2}}.e^{-\gamma /2}$\\
    \cline{1-4}
     $IV$& \footnotesize	$e^{\gamma /2}=\cosh \left(\frac{1}{2} \sqrt{4 a b+d^2}\right)+\frac{d \sinh
   \left(\frac{1}{2} \sqrt{4 a b+d^2}\right)}{\sqrt{4 a b+d^2}}$ & \footnotesize	$\alpha= \frac{\frac{2 a \sinh
   \left(\frac{1}{2} \sqrt{4 a b+d^2}\right)}{\sqrt{4 a b+d^2}}}{e^{-\gamma /2}}$&\footnotesize	 $\beta=\frac{2 b \sinh
   \left(\frac{1}{2} \sqrt{4 a b+d^2}\right)}{\sqrt{4 a b+d^2}}.e^{-\gamma /2}$\\
    \cline{1-4}
       $V$& \footnotesize	$e^{\gamma /2}=\cosh \left(\frac{1}{2} \sqrt{4 a b+d^2}\right)+\frac{d \sinh
   \left(\frac{1}{2} \sqrt{4 a b+d^2}\right)}{\sqrt{4 a b+d^2}}$& \footnotesize	$\alpha= \frac{2 a \sinh
   \left(\frac{1}{2} \sqrt{4 a b+d^2}\right)}{\sqrt{4 a b+d^2}}.e^{-\gamma /2}$ &\footnotesize	 $\beta=\frac{\frac{2 b \sinh
   \left(\frac{1}{2} \sqrt{4 a b+d^2}\right)}{\sqrt{4 a b+d^2}}}{e^{-\gamma /2}}$\\
    \cline{1-4}
       $VI$& \footnotesize	$e^{-\gamma /2}=\cosh \left(\frac{1}{2} \sqrt{4 a b+d^2}\right)-\frac{d \sinh
   \left(\frac{1}{2} \sqrt{4 a b+d^2}\right)}{\sqrt{4 a b+d^2}}$& $ \footnotesize	\alpha= \frac{2 a \sinh
   \left(\frac{1}{2} \sqrt{4 a b+d^2}\right)}{\sqrt{4 a b+d^2}}.e^{-\gamma /2}$ & \footnotesize	$\beta=\frac{\frac{2 b \sinh
   \left(\frac{1}{2} \sqrt{4 a b+d^2}\right)}{\sqrt{4 a b+d^2}}}{e^{-\gamma /2}}$\\
    \hline
\end{tabular}
\end{table}
\\
\newpage

\section{Colpa trick}
In this appendix we collect the main aspects of the Colpa trick which we used extensively in this paper. The relevant Hilbert space $\mathcal{H}$ for the operator $\mathcal{F}_{\mathbf{M},\bold{u},\bold{v}}$ has the dimension $2^L$. By adding the ancillary site, the corresponding Hilbert space $\mathcal{H}'$ of the new operator $\mathcal{F}_{\mathbf{M}'}$ will increase to $2^{L+1}$. The action of the operators in \eqref{eq: ancillary site} on the basis vectors of $\mathcal{H}'$ is presented in the table \ref{table7}.
\begin{table}[h]
		\caption{Action of operators on basis vectors of $\mathcal{H}'$. ($j=1,\cdots L$, $n_j=0,1$ and $m=n_1+n_2+\cdots + n_{j-1}$).\label{table7}}
\scriptsize	\begin{tabular}{SSSSS} \toprule
		{} & {$\ket{0 n_1 n_2 \cdots n_j=0 \cdots n_L}$} & {$\ket{1 n_1 n_2 \cdots n_j=0 \cdots n_L}$} & {$
		\ket{0 n_1 n_2 \cdots n_j=1 \cdots n_L}$} & {$\ket{1 n_1 n_2 \cdots n_j=1 \cdots n_L}$}  \\ \midrule
		{$c_{0}^{\dagger}c_{j}-c_{0}c_{j}$}  & {$\ket{\mathbf{0}}$} & {$\ket{\mathbf{0}}$} & {$(-1)^m \ket{1 n_1 n_2 \cdots n_j=0 \cdots n_L}$} & {$(-1)^m \ket{0 n_1 n_2 \cdots n_j=0 \cdots n_L}$} \\
		{$c_j^{\dagger}c_0-c_j^{\dagger}c_0^{\dagger}$}  & {$(-1)^m \ket{1 n_1 n_2 \cdots n_j=1 \cdots n_L}$}  & {$(-1)^m \ket{0 n_1 n_2 \cdots n_j=1 \cdots n_L}$} & {$\ket{\mathbf{0}}$}  & {$\ket{\mathbf{0}}$}\\ \bottomrule
	\end{tabular}
		\end{table}
\newpage
Using the above table one can define the projected Hilbert space $\mathcal{H}'_+$ spanned by the the following basis:

\begin{equation}\label{eq: orthonormal set}
    \frac{1}{\sqrt{2}}(|0n_1...n_L\rangle+|1n_1...n_L\rangle).
\end{equation} 
The main idea of Colpa trick is that the action of $\mathcal{F}_{\mathbf{M},\bold{u},\bold{v}}$ on $\mathcal{H}$ is the same as the action of 
$\mathcal{F}_{\mathbf{M}^{\prime}}$ on $\mathcal{H}'_+$. Note that in the language of the spins this is like projecting the ancillary spin in the up direction in the $\sigma^x$ basis.

\section*{References}
\providecommand{\newblock}{}

\end{document}